\documentclass[a4paper,fleqn,usenatbib,useAMS]{mnras}

\usepackage{newtxtext,newtxmath}

\usepackage[T1]{fontenc}
\usepackage{ae,aecompl}

\usepackage{graphicx}	
\usepackage{amsmath}	
\numberwithin{equation}{section} 
\usepackage{amssymb}	
\usepackage[separate-uncertainty = true,multi-part-units=brackets]{siunitx}      
\usepackage{wasysym}      
\usepackage{commath}      
\usepackage{physics}  
\usepackage[noabbrev]{cleveref}     
\usepackage[final]{changes} 
\makeatletter
\@namedef{Changes@AuthorColor}{red}
\colorlet{Changes@Color}{red}
\makeatother

\newcommand*\mean[1]{\langle #1 \rangle}
\newcommand{\HII}{\ion{H}{II}} 

\newcommand{\ue}{\textrm{e}} 

\crefformat{equation}{\eqA equation~\eqB #2#1#3)}
\newcommand{\eqA}{}
\newcommand{\eqB}{(}

\crefname{figure}{Fig.}{Figs.}
\Crefname{figure}{Fig.}{Figs.}
\crefname{table}{Table}{Tables}
\Crefname{equation}{Equation}{Equations}
\newcommand{\solar}{\ensuremath{_{\astrosun}}} 
\DeclareSIUnit\erg{erg}
\DeclareSIUnit\dyne{dyne}
\DeclareSIUnit\msol{M\solar{}}
\DeclareSIUnit\lsol{L\solar{}}
\DeclareSIUnit\rsol{R\solar{}}
\DeclareSIUnit\yr{yr}
\DeclareSIUnit\pc{pc}
\DeclareSIUnit\str{str}
\DeclareSIUnit\jy{Jy}
\DeclareSIUnit\micron{\micro\metre}

\title[Variation of the FUV ISRF]{Massive star feedback in clusters: variation of the FUV interstellar radiation field in time and space}

\author[A. A. Ali and T. J. Harries]{
Ahmad A. Ali\thanks{E-mail: aali@astro.ex.ac.uk} and
Tim J. Harries 
\\
Department of Physics and Astronomy, University of Exeter, Stocker Road, Exeter EX4 4QL, United Kingdom
}

\date{Accepted ???. Received ???; in original form ???}

\pubyear{2018}

\begin{document}
\label{firstpage}
\pagerange{\pageref{firstpage}--\pageref{lastpage}}
\maketitle

\begin{abstract}
We investigate radiative feedback from a \SI{34}{\msol} star in a \SI{e4}{\msol} turbulent cloud using three-dimensional radiation-hydrodynamics (RHD) models. We use Monte Carlo radiative transfer to accurately compute photoionization equilibrium and radiation pressure, with multiple atomic species and silicate dust grains. We include the diffuse radiation field, dust absorption/re-emission, and scattering. 
The cloud is efficiently dispersed, with 75 per cent of the mass leaving the $(\SI{32.3}{\pc})^3$ grid within \SI{4.3}{\mega\yr} (1.1 $\mean{t_\textrm{ff}}$). This compares to all mass exiting within \SI{1.6}{\mega\yr} (0.74 $\mean{t_\textrm{ff}}$) in our previously published \SI{e3}{\msol} cloud. At most 20 per cent of the mass is ionized, compared to 40 per cent in the lower mass model, despite the ionized volume fraction being 80 per cent in both, implying the higher mass cloud is more resilient to feedback. The total Jeans-unstable mass increases linearly up to \SI{1500}{\msol} before plateauing after \SI{2}{\mega\yr}, corresponding to a core formation efficiency of 15 per cent.
We also measure the time-variation of the far-ultraviolet (FUV) radiation field, $G_0$, impinging on other cluster members, taking into account for the first time how this changes in a dynamic cluster environment with intervening opacity sources and stellar motions. Many objects remain shielded in the first \SI{0.5}{\mega\yr} whilst the massive star is embedded, after which $G_0$ increases by orders of magnitude. Gas motions later on cause comparable drops which happen instantaneously and last for $\sim \SI{1}{\mega\yr}$ before being restored. This highly variable UV field will influence the photoevaporation of protoplanetary discs near massive stars.

\end{abstract}




\section{Introduction}
\defcitealias{ali2018}{Paper I}
Star formation occurs in giant molecular clouds (GMCs) ranging in mass from $\sim 10^4$ to \SI{e6}{\msol}, with radii of a few to \SI{100}{pc} \citep{solomon1987,heyer2009}. The dynamics in these GMCs are likely dominated by massive O stars via radiative feedback \citep{matzner2002} occurring in two wavelength regimes: extreme-ultraviolet (EUV) photons with energy greater than \SI{13.6}{\eV} can ionize atomic hydrogen to form an \HII{} region. Far-ultraviolet (FUV) radiation between 5 and \SI{13.6}{\eV}, to which ionized gas is optically thin, is absorbed by molecules, leading to them being photodissociated into their constituent atoms \citep{stecher1967}. Furthermore, FUV absorption by dust grains can heat the grains themselves or heat the gas via photoelectric ejection  \citep{draine1978,wolfire1995,hollenbach1999}. Massive stars are therefore surrounded by an \HII{} region of ionized gas at \SI{e4}{K}, bounded by a photodissociation region (PDR) of atomic gas between 100 to \SI{1000}{K}. This is finally followed by molecular gas at \SI{10}{K}.

The thermal pressure gradient between the warm ionized gas and the cold neutral surroundings causes the \HII{} region to expand into the neutral medium. The dispersal of gas via expanding \HII{} regions has been well studied using analytical and numerical models,   including two-dimensional studies of champagne flows \citep{whitworth1979,tenorio-tagle1979}, filamentary configurations \citep{bodenheimer1979}, and power-law density fields \citep{franco1990}. These studies showed how ionizing feedback from massive stars can efficiently disperse their host molecular clouds, even with a small star formation efficiency (SFE) of a few per cent, as is observed in the Galaxy \citep{lada2003}.  However, the problem becomes more complex in three-dimensional, inhomogeneous gas distributions, as shown by \citet{dale2005}; unlike previous works, they concluded that clouds may not necessarily be destroyed by ionizing radiation, as a small fraction of the gas may effectively shield the rest of the cloud, carrying significant amounts of energy out of the system instead of distributing it evenly. Parameter studies by \citet{dale2011} and \citet{dale2012,dale2013a} found that the degree of dispersal was closely coupled to the initial conditions of their simulations, such as cloud mass \citep[see also][]{howard2017a}. Dispersal has also been shown to depend on initial gas surface density \citep{kim2018a}, morphology \citep{geen2018}, and cluster luminosity \citep{geen2016,geen2018}. One problem is that even the most recent models simplify the radiative transfer, for example by using the on-the-spot approximation for recombination (thus neglecting the ionizing photons re-emitted by the gas), by using simple two-step temperature schemes for neutral and ionized hydrogen, or by neglecting dust microphysics such as absorption and scattering. These have a non-negligible effect on dynamics and morphology \citep[see e.g.][]{ercolano2011,haworth2012,haworth2015} and therefore must be accounted for to accurately model gas dispersal by radiative feedback.

In addition to dispersal on a grand scale, there is also the thermodynamical evolution of individual structures inside the \HII{} region. Proplyds -- neutral clumps of density $\sim\SI{e5}{\per\cm\cubed}$, size $\sim$ 100 to \SI{e5}{AU}, with ionized boundaries, and containing a circumstellar or protoplanetary disc -- are seen in the vicinity of O stars, most notably near $\theta^1$ Ori C in the Orion Nebula Cluster  \citep{odell1993,bally2000,odell2001}. They have also been seen in other \HII{} regions such as NGC 3603 \citep{brandner2000}, the Carina Nebula \citep{smith2003}, Cygnus OB2 \citep{wright2012}, and even around B stars with weaker UV fields \citep{kim2016a}. EUV photons ionize the outermost layer of a proplyd, resulting in a photoevaporative flow of ionized gas into the diffuse environment. FUV photons are able to penetrate further in, photodissociating and heating the molecular disc surface, which drives a neutral atomic wind from the PDR (as modelled by \citealt{johnstone1998} and measured spectroscopically by \citealt{henney1999}). 

Many observational studies of different star-forming regions have found a positive correlation between disc frequency/size/mass and projected distance from ionizing sources \citep{odell1998,guarcello2007,guarcello2010,guarcello2016,ansdell2017,eisner2018}.  Hydrodynamic models indeed show a correlation between extent of disc photoevaporation and $G_0$ \citep{adams2004,clarke2007,anderson2013,haworth2017,winter2018,haworth2018c}, where $G_0$ is the FUV flux in units of the \citet{habing1968} field (\SI{1.63e-3}{\erg\per\s\per\cm\squared}). Combined with simulations which show that disc destruction may occur primarily through external photoevaporation rather than stellar encounters \citep{scally2001,winter2018}, the interstellar radiation field (ISRF) appears to be of crucial importance in determining the evolution of disc properties. The above studies explore the parameter space of $G_0$ from a solar neighbourhood value of 1 to an inner ONC value of $10^6$, depending on an assumed distance to an FUV source. Real proplyds and discs are located within gaseous, dusty, and turbulent clouds. Massive stars disperse material, as described above, and this alters the column density between them and any proplyds. Furthermore, the cluster members have their own motions, meaning the separations also change. Stellar velocity dispersions are of the order \SI{1}{\kilo\m\per\s} ($=\SI{1}{\pc\per\mega\yr}$), proplyds can be propelled to \SI{5}{\kilo\m\per\s}, and the ionized sound speed is \SI{10}{\kilo\m\per\s} -- therefore, over the course of a proplyd lifetime, there will be non-negligible changes in the incident flux due to the kinematics of both the stars and the gas; both effects combined will vary the geometric dilution ($\propto r^{-2}$ for separation $r$) and the FUV attenuation ($\propto \ue^{-\tau}$ for optical depth $\tau$) as a function of time and distance. This is the behaviour we aim to describe in this paper.

In \citet[hereafter Paper I]{ali2018}, we modelled a \SI{e3}{\msol} cloud with a \SI{34}{\msol} massive star, using a detailed radiative transfer scheme to compute photoionization and radiation pressure -- this included the diffuse radiation field, scattering and absorption by dust, and calculations of thermal equilibrium with cooling from recombination lines, forbidden lines, and free-free continuum. In this paper, we apply the same scheme to a 10-times more massive cloud, using the same initial mean surface density and stellar luminosity. We outline the numerical methods in \cref{sec:numericalmethods} and the initial setup in \cref{sec:m1e4-initialconditions}. We present and discuss the results in \cref{sec:feedback} (for the dynamics) and \cref{sec:m1e4-fuvresults} (for the FUV ISRF). Finally, we draw our conclusions in \cref{sec:conclusions}.

\section{Numerical methods}
\label{sec:numericalmethods}
We use the Monte Carlo (MC) radiative transfer (RT) and hydrodynamics (HD) code, \textsc{torus} \citep[described in detail by][]{harries2019}. We use the same treatment as in \citetalias{ali2018}, but provide a summary here. The MCRT technique follows the \citet{lucy1999} method, whereby energy packets representing photons are propagated through a medium and undergo scattering and absorption/re-emission events. This includes the radiation field from stars as well as the diffuse field from gas and dust. Photon wavelengths are interpolated from 1000 logarithmically spaced bins between \SI{e2}{} and \SI{e7}{\angstrom}. Silicate dust grains \citep{draine1984} are distributed according to a standard ISM power-law density relationship \citep{mathis1977}
\begin{equation}
	\label{eq:dustdistribution}
    n(a) \propto a^{-q}
\end{equation}
using grain sizes $a$ between 0.1 to \SI{1}{\micron} and a power-law index $q$ of 3.5, giving a median grain size of \SI{0.12}{\micron}. The dust-to-gas mass ratio is kept constant at 0.01. The dust temperature is calculated from the balance between photon absorption and thermal Planck emission. This is calculated separately from the gas temperature, but an additional term accounts for collisional heat exchange between the two \citep{hollenbach1979}.  For the gas, we calculate photoionization balance for \ion{H}{i--ii}, \ion{He}{i--iii}, \ion{C}{i--iv}, \ion{N}{i--iii}, \ion{O}{i--iii}, \ion{Ne}{i--iii}, and \ion{S}{i--iv}. Thermal equilibrium for the gas temperature takes into account heating due to photoionization of H and He, and cooling from H and He recombination lines, collisionally excited metal forbidden lines, and free--free continuum \citep{haworth2012}.

Radiative transfer is calculated before every hydrodynamics step, which evolves isothermally. Self-gravity is included, using a V-cycling multigrid method to solve Poisson's equation with Dirichlet boundary conditions based on a multipole expansion. Radiation pressure is calculated using the momentum-transfer algorithm presented in \citet{harries2015}. Sink particles \citep[see][]{harries2015} are included but do not accrete material or form on-the-fly (see \cref{sec:m1e4-initialconditions} for the formation method); however, they do evolve in mass, effective temperature, and luminosity using \citet{schaller1992} evolutionary tracks. O stars emit radiation according to \textsc{ostar2002} \citep{lanz2003} spectra, while later-type stars use \citet{kurucz1991} spectra.  They move due to N-body interactions and the mutual gravitational force from gas. These are used as tracer particles to follow the behaviour near the possible locations of actual stars or proplyds.

The FUV radiation field is calculated in each cell as part of the MCRT algorithm, using
\begin{equation}
	\label{eq:g0}
	\begin{split}
	G_0 &=  \frac{1}{H} \int_{\SI{912}{\angstrom}}^{\SI{2400}{\angstrom}} 4 \upi J_{\lambda} \dif \lambda \\
	&= \frac{1}{H} \frac{\epsilon}{\Delta t V} \sum_{\lambda=\SI{912}{\angstrom}}^{\SI{2400}{\angstrom}}  \ell 
	\end{split}
\end{equation}
given in units of the \citet{habing1968} field, $H = \SI{1.63e-3}{\erg\per\s\per\cm\squared}$. $J_\lambda$ is the mean intensity, $V$ is the cell volume, and $\ell$ is the path length travelled between events (scattering/absorption/cell-boundary crossing) by a MC photon packet with energy $\epsilon$ during time step $\Delta t$. This is calculated in every cell at all time steps. To track how it can vary at the location of a particular sink particle (representing a star or proplyd), we follow $G_0$ in the cell containing that sink. When a sink moves from one cell to another, the intensity may jump up or down due to the discrete nature of the grid structure (and hence the optical depth). Therefore, in the results presented in \cref{sec:m1e4-fuvresults}, we average around a radius of 2.5 cells; if accretion were enabled, this would represent the computational accretion radius, but is simply a convenience factor here. For \cref{sec:m1e4-fuvresults}, we integrate the optical depth along the ray joining two sinks as a measure of the dust attenuation; that is,
\begin{equation}
	\label{eq:taufuv}
	\tau_\textrm{FUV} = \int \kappa_\textrm{FUV} \rho \dif s
\end{equation}
where $\rho$ is density and $\kappa_\textrm{FUV}$ is a representative average of the dust opacity in each cell given by
\begin{equation}
	\label{eq:meanopacity}
    \kappa_\textrm{FUV} = \frac{ \int_{\SI{912}{\angstrom}}^{\SI{2400}{\angstrom}} (\kappa_\textrm{abs} + \kappa_\textrm{sca}) \dif \lambda }
    {\int_{\SI{912}{\angstrom}}^{\SI{2400}{\angstrom}} \dif \lambda  }
\end{equation}
for absorption and scattering opacities $\kappa_\textrm{abs}$ and $\kappa_\textrm{sca}$, respectively. It should be noted that the radiative transfer uses the actual opacities, not this average value.

\section{Initial conditions}
\label{sec:m1e4-initialconditions}

We model a \SI{e4}{\msol} cloud which is initially spherical with a mean surface density of \SI{0.01}{\g\per\cm\squared} and radius $R_s = \SI{8.41}{\pc}$. The mass volume density is uniform in the inner core  (\SI{4.9e-22}{\g\per\cm\cubed} between $0 > r > R_s/2$), then falls off as $r^{-1.5}$ (reaching \SI{1.7e-22}{\g\per\cm\cubed} at $r = R_s$). The density outside the cloud is 1 per cent of the density at the sphere boundary. These conditions are similar to \citetalias{ali2018}, where the sphere mass and radius were \SI{e3}{\msol} and \SI{2.66}{\pc}, respectively. The mean surface densities of both spheres are the same. We refer to the \SI{e4}{\msol} cloud presented in this paper as the `high-mass cloud', and the  \SI{e3}{\msol} cloud in \citetalias{ali2018} as the `low-mass cloud'.

The free-fall time associated with a uniform sphere with density $3M_s/4 \pi R_s^3 = \SI{3e-22}{\g\per\cm\cubed}$ is $\mean{t_\textrm{ff}} = \SI{3.86}{\mega\yr}$ (where $M_\textrm{s}$ is the total sphere mass). The temperature is \SI{10}{K} everywhere for both gas and dust until radiation sources are switched on. The grid size from end to end is \SI{32.3}{\pc}, approximately 4 times the sphere radius, giving a linear resolution of \SI{0.13}{\pc} per cell with $256^3$ cells. The grid structure is Cartesian, uniform, and fixed. 

\begin{figure}
    \centering
	\includegraphics[width=\columnwidth]{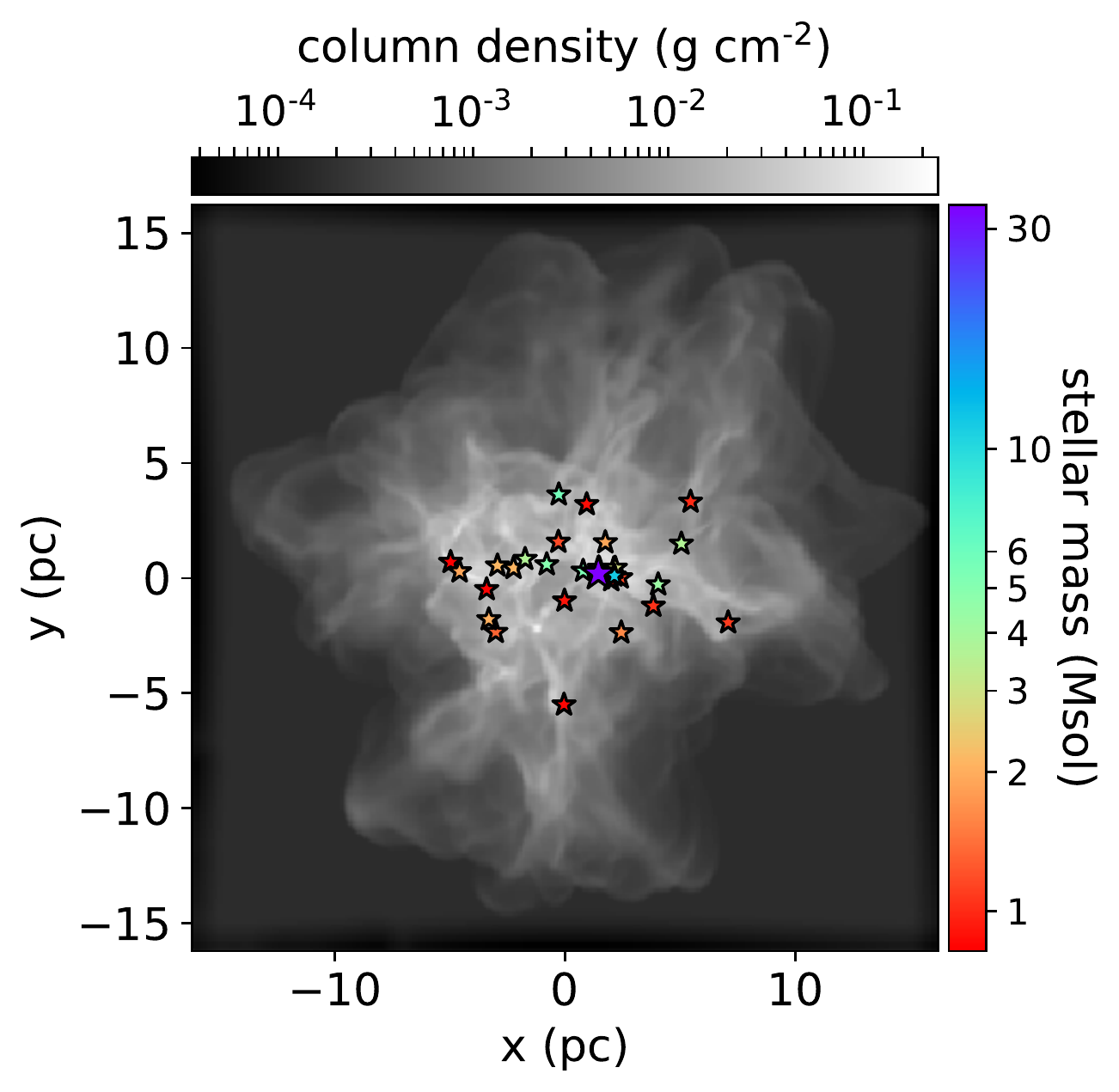}
    \caption{Stellar masses (colour scale) and positions in projection when feedback is switched on. Gas column density is shown in greyscale.}
    \label{fig:m1e4-starburst}
\end{figure}

\begin{figure}
    \centering
	\includegraphics[width=\columnwidth]{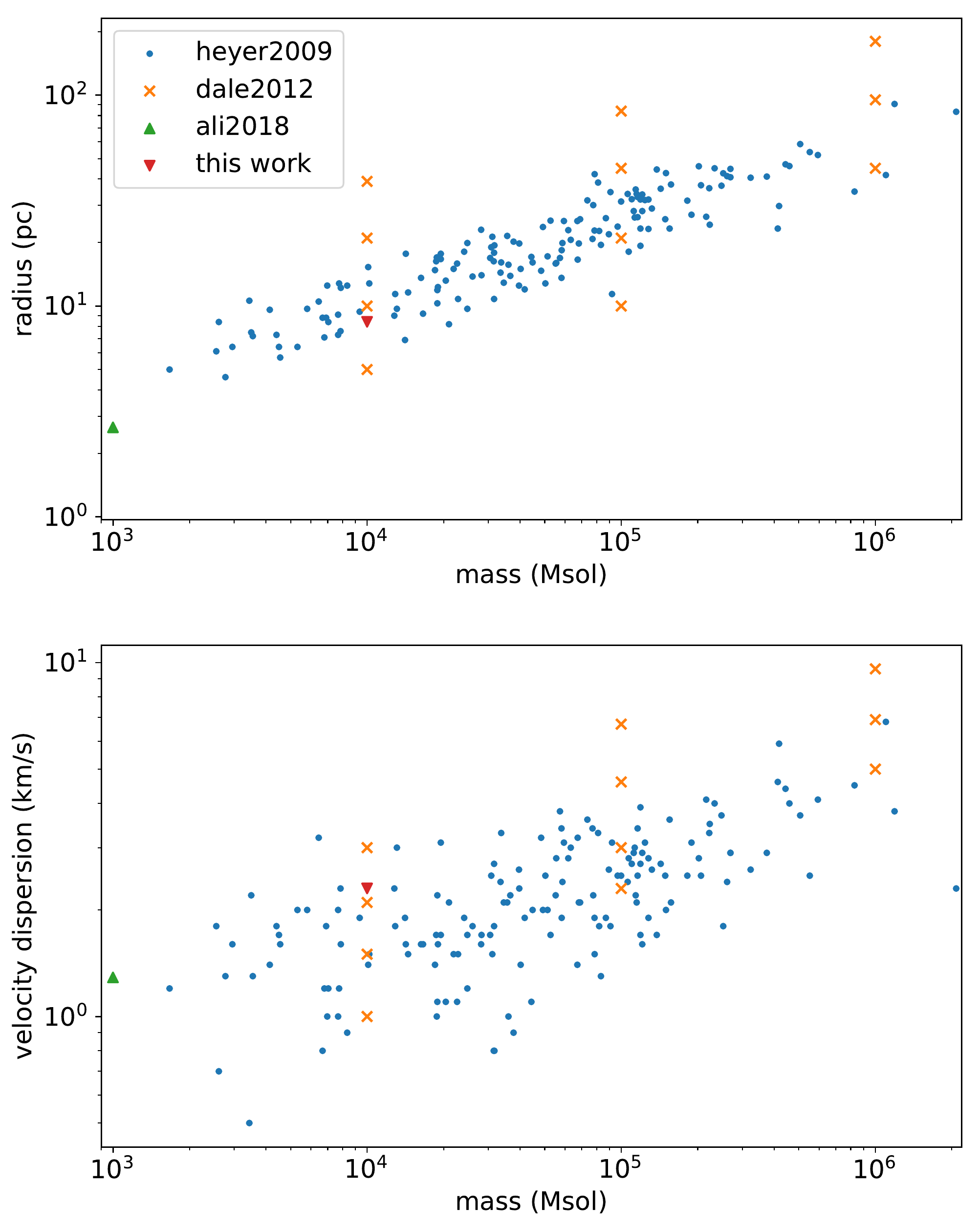}
    \caption[Parameter space of observed and simulated clouds]{Parameter space of Galactic GMC observations by \citet[dots]{heyer2009}, simulations by \citet[crosses]{dale2012}, \citet[up-pointing triangle]{ali2018}, and this work (down-pointing triangle).}
    \label{fig:parameterspace}
\end{figure}

As in \citetalias{ali2018}, the sphere evolves under self-gravity and a seeded turbulent velocity field for 0.75\,$\mean{t_\textrm{ff}}$ without stars. We use the same random Gaussian turbulent velocity field as \citet{bate2002}, with a power spectrum $P(k)\propto k^{-4}$ for wavenumber $k$, such that the kinetic energy equals the gravitational potential energy, i.e. the virial parameter $\alpha_\textrm{vir} \equiv 2 E_\textrm{kin}/E_\textrm{grav} = 2$. At 0.75\,$\mean{t_\textrm{ff}}$, a \SI{33.7}{\msol} star is placed in the cloud's most massive core. 28 other stars are inserted according to a probability density function (PDF) proportional to star formation rate $\dot{M}(\mathbfit{r})$ at some position $\mathbfit{r}$; that is, $p(\mathbfit{r}) \propto \dot{M}(\mathbfit{r}) \propto \rho(\mathbfit{r})/t_\textrm{ff} \propto \rho(\mathbfit{r})^{1.5}$. We keep the same number of stars and stellar masses as \citetalias{ali2018} in order to compare how the same luminosity affects a more massive cloud. This means the imposed star formation efficiency in the high-mass model is 1 per cent as opposed to 10 per cent in the low-mass model. The initial velocity of each star is the velocity of the gas in the cell containing that star. The initial positions (plotted in projection in \cref{fig:m1e4-starburst}) are not necessarily the same as in \citetalias{ali2018}, as they are placed using the aforementioned PDF without any further constraints; however, since the turbulence in the initial hydrodynamics-only phase results in a similar morphology at 0.75\,$\mean{t_\textrm{ff}}$, when stars are created, the most massive star ends up being placed in roughly the same area of the cloud. This keeps the initial conditions broadly consistent, but just scaled up in cloud mass. The parameter space of cloud mass, radius, and velocity dispersion is shown in \cref{fig:parameterspace}, alongside observations of GMCs in the Galaxy by \citet{heyer2009} and the suite of simulations by \citet{dale2012}.


\section{Dynamics}
\label{sec:feedback}
In this section we present and discuss the results of the RHD simulations for the \SI{e4}{\msol} cloud. One includes photoionization and radiation pressure, and one has photoionization only. The models are evolved for  $\sim$ \SI{4}{\mega\yr} (1\,$\mean{t_\textrm{ff}}$), with $t=0$ corresponding to the initiation of feedback. Results for the \SI{e3}{\msol} model (`low-mass cloud') are presented in \citetalias{ali2018}; this was evolved under feedback for \SI{1.6}{\mega\yr} (0.74\,$\mean{t_\textrm{ff}}$).

\subsection{Bulk grid properties}
\label{sec:m1e4-bulkgridproperties}

\begin{figure*}
    \centering
	\includegraphics[width=0.95\textwidth]{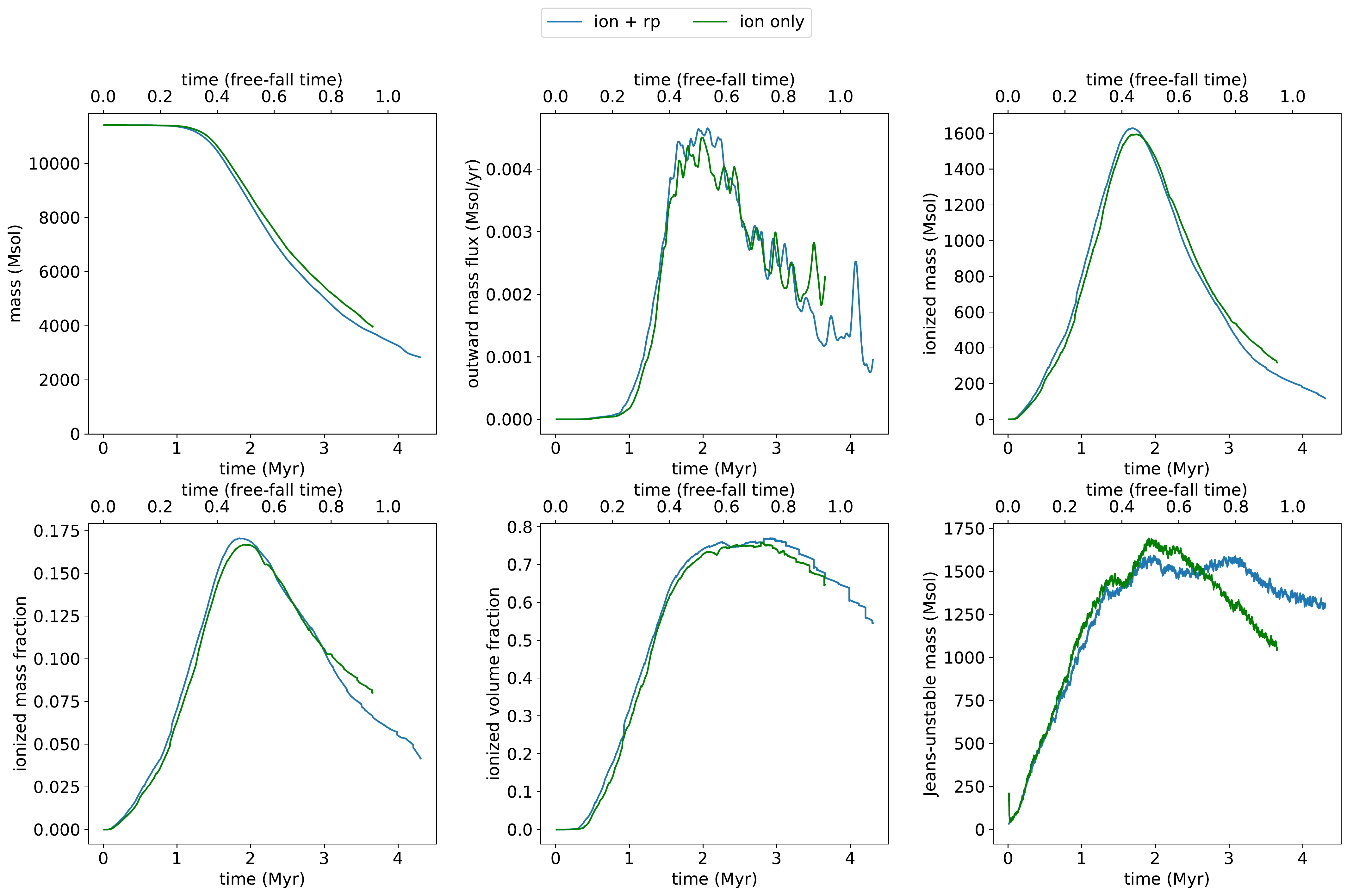}
    \caption{Time-varying grid properties for the \SI{e4}{\msol} cloud model showing the total mass on grid, mass flux off the grid, ionized mass, ionized mass fraction, ionized volume fraction, and total mass in Jeans-unstable cells. The blue line includes ionization and radiation pressure, while the green line only includes ionization. Feedback starts at $t=0$.}
    \label{fig:m1e4-plotoftotals}
\end{figure*}

\Cref{fig:m1e4-plotoftotals} shows the grid properties as a function of time. The first panel plots the total mass, which starts just under \SI{1.2e4}{\msol} (including the low-density material outside the \SI{e4}{\msol} turbulent sphere); between 1 and \SI{4}{\mega\yr}, there is a steady decrease as mass leaves the grid. The total mass flux at the grid boundaries, shown in the second panel of \cref{fig:m1e4-plotoftotals}, peaks between 1.5 and \SI{2.5}{\mega\yr} (0.39 and 0.65 $\mean{t_\textrm{ff}}$ respectively) with a value of \SI{4.7e-3}{\msol\per\yr}. This is more than double the mass flux in the \SI{1000}{\msol} model presented in \citetalias{ali2018}. The shape of the curve is essentially the same, with the first phase being dominated by the removal of low-density material, peaking at the crossing time for ionized gas travelling from the centre of the grid to the edge, with a sound speed $\approx \SI{10}{\kilo\m\per\s}$; the decreasing second phase is overlaid with sharp, short-lived spikes corresponding to the removal of dense clumps. 

At \SI{1.75}{\mega\yr} (0.45 $\mean{t_\textrm{ff}}$), the highest value of ionized mass is reached, with \SI{1600}{\msol}, or about 17 per cent, of the mass being ionized; this proportion is half that of the low-mass model, which peaked at 40 per cent, again at the time of peak mass flux. This is despite about 80 per cent of the volume being ionized, which is comparable to the low-mass model. The conclusion reached in that model is even more so the case here -- most of the mass remains in small, dense, neutral areas.

The final frame of \cref{fig:m1e4-plotoftotals} shows the total mass in Jeans-unstable cells.It should be noted that dust-heating by (F)UV absorption raises the temperature from \SI{10}{K} to a high of $\sim$ \SI{100}{K}, which results in a higher Jeans mass and therefore greater resistance against fragmentation. The figure represents an upper limit to the amount of mass that could fragment to form stars. Only a third of a core mass may actually be converted into stars \citep{alves2007}. Furthermore, to prevent artificial numerical fragmentation, the Jeans length
\begin{equation}
    \label{eq:jeanslength}
    \lambda_J = \sqrt{\frac{\upi c_s^2}{G\rho}}
\end{equation}
should be resolved by at least four grid cells \citep{truelove1997}. This limit corresponds to a density $\sim \SI{e-20}{\g\per\cm\cubed}$. 
 
The total Jeans-unstable mass increases linearly to \SI{1500}{\msol} before plateauing at around \SI{2}{\mega\yr} (0.5 $\mean{t_\textrm{ff}}$). Dense clumps start leaving the grid after this stage, which results in the unstable mass decreasing. In terms of the original cloud mass, this represents a core formation efficiency of 15 per cent, or a star formation efficiency of 5 per cent. This order of magnitude is in line with observations in the Galaxy \citep{lada2003}. It also agrees with the simulation by \citet{geen2017}, labelled `L', which has a similar initial mass and surface density, and includes ionization feedback -- \citet{geen2017} found that the SFE for that cloud was closer to observations of local star-forming regions compared to denser models, where SFEs approached 100 per cent. 

The bulk grid properties for the model without radiation pressure are in general similar to the combined feedback model, although there are minor differences. For example, there is a difference of \SI{300}{\msol} for the total Jeans-unstable mass. Since the inferred core- or star-formation efficiency is a rough estimate, this difference may not be significant, but indicates the possibility of radiation pressure producing denser structures. 

\begin{figure*}
    \centering
	\includegraphics[width=\textwidth]{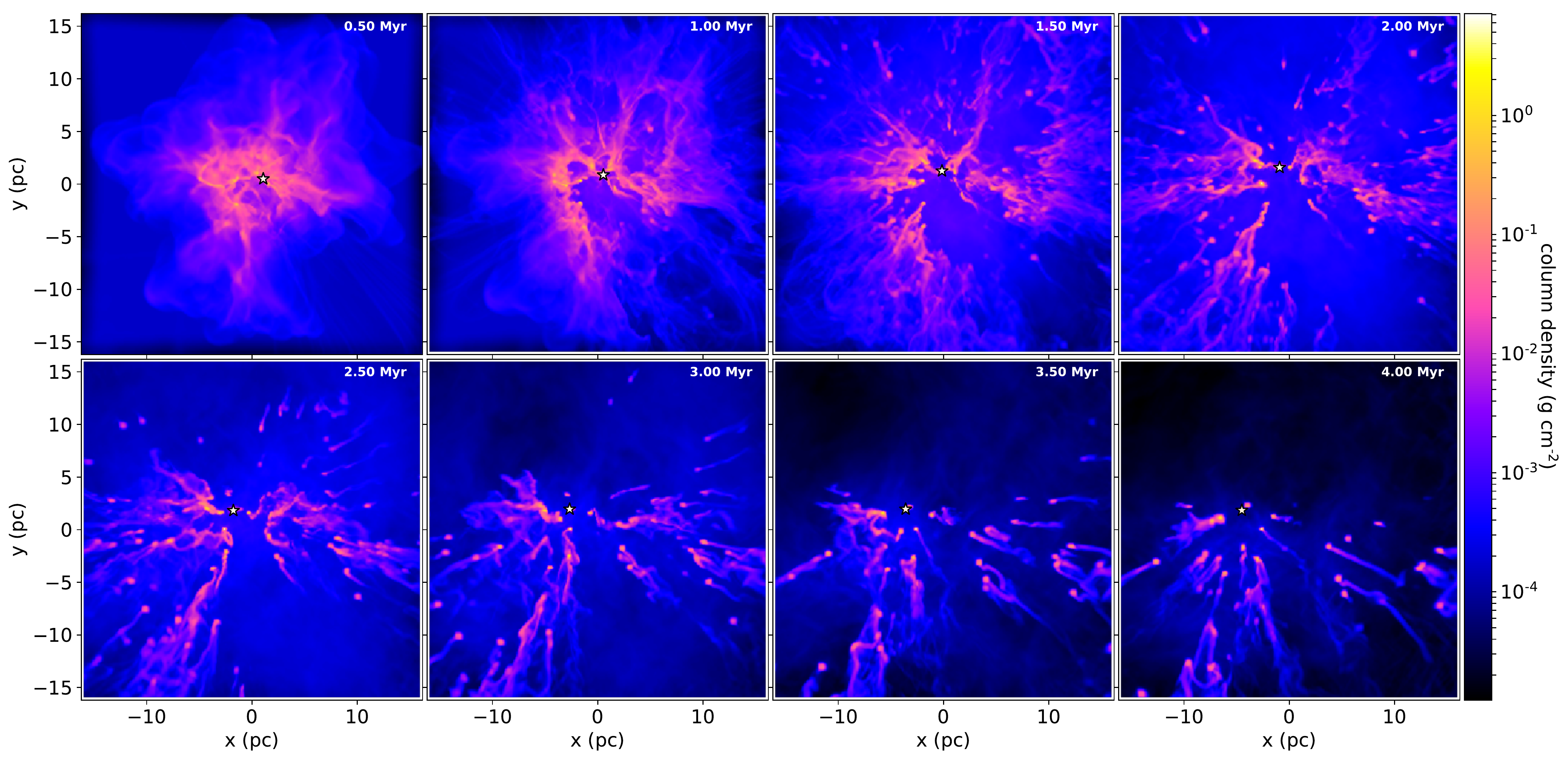}
    \caption{Column density in the $x$-$y$ plane at 0.5 Myr intervals for the model with both photoionization and radiation pressure. The \SI{34}{\msol} star is shown as a white point. Feedback starts at $t=0$.}
    \label{fig:m1e4-columndensity}
\end{figure*}

\begin{figure}
    \centering
	\includegraphics[width=\columnwidth]{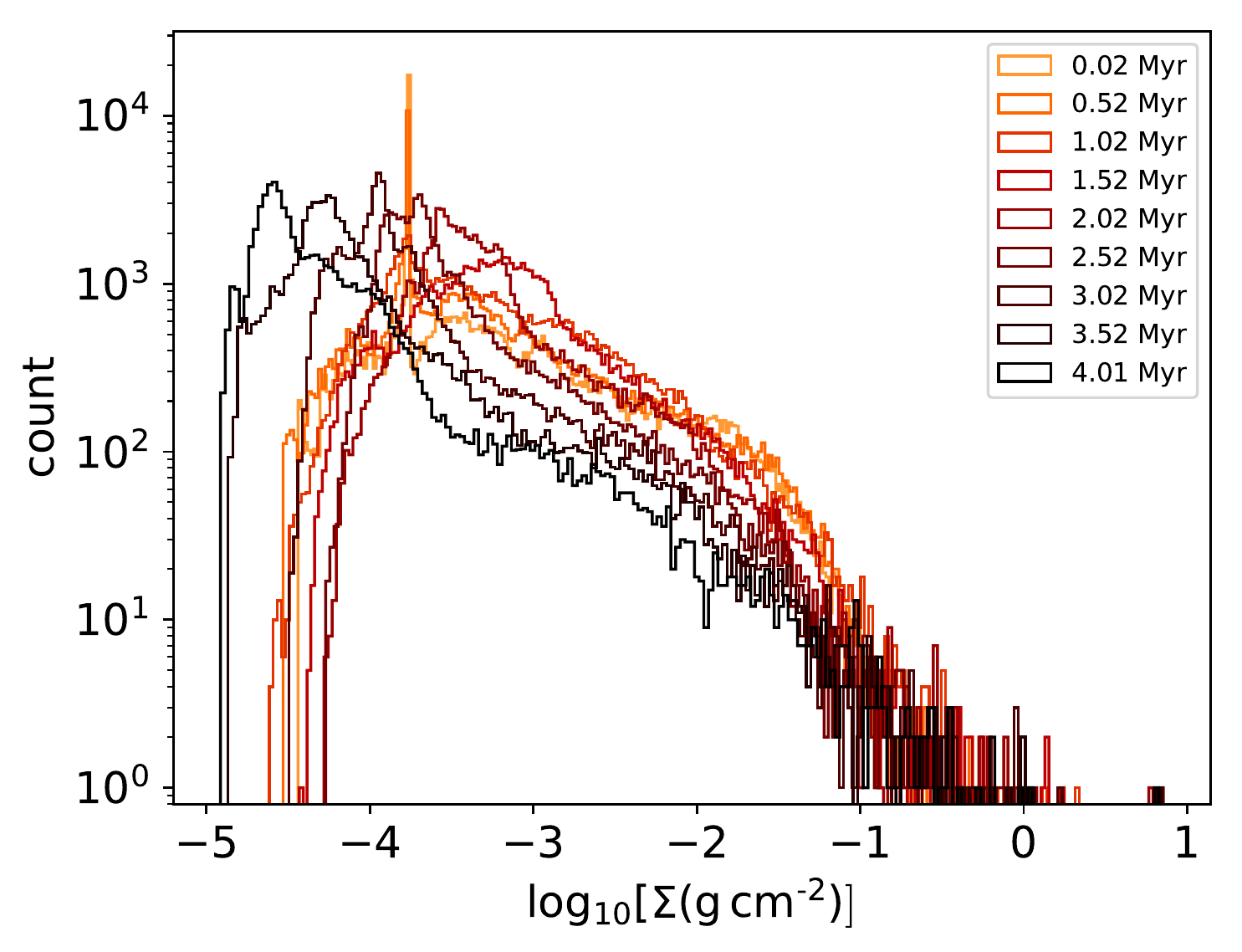}
    \caption{Column density histograms at 0.5 Myr intervals for the model with both feedback mechanisms. The spike at $t=0$ is caused by the uniform-density gas outside the sphere.}
    \label{fig:m1e4-densitypdf}
\end{figure}

\Cref{fig:m1e4-densitypdf} shows column density histograms for the first \SI{4}{\mega\yr}. Compared to the low-mass cloud in \citetalias{ali2018}, they have similar shapes but are more homogeneous with time, only significantly shifting to lower surface densities beyond \SI{3}{\mega\yr}. This model starts off with more material in the low-density end, at $\Sigma = \SI{e-4.5}{\g\per\cm\squared}$, whereas the low-mass cloud stops at $\Sigma = \SI{e-4}{\g\per\cm\squared}$; therefore this model requires less of a left-ward shift to reach the $\Sigma = \SI{e-5}{\g\per\cm\squared}$ peak by the end of the simulation. There is very little variation during the first \SI{1.5}{\mega\yr} (the total run time for the low-mass model), other than the spreading out of the low-density spike ($\sim \SI{e-4}{\g\per\cm\squared}$) corresponding to the diffuse gas outside the cloud. 

\subsection{Dispersal}
As in \citetalias{ali2018}, the dense filaments on the left side of the cloud are more resistant to being dispersed than the right side. In particular, a cavity is easily blown out in the bottom right quadrant of the $x$-$y$ plane in the first \si{\mega\yr}, allowing ionized material to stream out. This is also the case at $(x=0, y>0)$ and $(x \approx \SI{3}{\pc},y>0)$. It is not surprising that these locations are where the \HII{} breaks out, as this also happens in the low-mass cloud -- both have a similar gravo-turbulent structure and a massive star in similar locations. 

However, in this massive cloud, the dense filaments on the left side ($x<0$ in the $x$-$y$ plane) have more time to accumulate mass before the \HII{} region can reach them. Indeed, the filaments are still collapsing towards the centre while feedback progresses. This makes them less prone to dispersal, whether by ionization heating or the rocket effect -- by $t=\SI{4}{\mega\yr}$, many of the globule--pillar objects are still present inside a \SI{8}{\pc} radius around the massive star, which corresponds to the radius of the entire grid in \citetalias{ali2018}. 

The initial conditions are similar to run I of \citet{dale2012} (the closest point to this model in \cref{fig:parameterspace}). Although their model starts off with a smoother density distribution than our model, a similar butterfly-shaped \HII{} region is produced -- this is the only model in their suite of conditions to create this morphology. Run I showed the greatest effect in terms of structure and dynamics, with 60 per cent of the gas being unbound. Their total ionization fraction did not exceed 10 per cent, whereas the model here approaches 20 per cent, even with similar ionizing fluxes ($\sim \SI{e49}{\per\s}$).

\citet{dale2013a} summarised their parameter study of GMC masses between $10^4$ to \SI{e6}{\msol}, stating that the ionized mass fraction is around 5 to 10 percent regardless of the cloud properties. This disagrees with the models presented here and in \citetalias{ali2018}, which approach fractions of 20 per cent and 40 per cent, respectively; that is, they are both higher and they vary with cloud mass. 
This is despite their models having ionizing luminosities which increase with cloud mass.  One contributor to the difference between models may be the diffuse radiation field, which the Dale et al. models not include. As shown by \citet{ercolano2011} and \citet{haworth2012}, this can penetrate into what would otherwise be shadowed regions behind dense gas, creating changes in morphology such as detaching pillar heads. If this makes the environment more permeable to ionizing photons, the total ionized mass could be increased. This bears more investigation as the method for forming stars differs between the two studies, as does the gas distribution when stars begin to radiate, and these differences can also change the effectiveness of feedback \citep{geen2018}. \citet{dale2013a} also find a strong dependence on dispersal efficiency with cloud escape velocity. $v_\textrm{esc}$ for the cloud presented here is still very much below the ionized sound speed ($\sim \SI{3}{\kilo\m\per\s}$ compared to \SI{10}{\kilo\m\per\s}, respectively), but the full parameter space is yet to be explored.

The ineffectiveness of radiation pressure as a dispersive feedback mechanism is in agreement with the analytical model of \citet{fall2010}, who showed that a \SI{e4}{\msol} cloud requires at least ten times greater surface density than the cloud we model here for radiation pressure to become dominant.  In their numerical models, \citet{kim2018a} found that surface density was the important parameter in determining the feedback efficiency; photoionization dominated over radiation pressure for similar initial conditions to ours, and in general the latter only became relevant once the surface density exceeded a few times that of our cloud. \citet{skinner2015} showed that the radiation pressure from dust-processed infrared photons was ineffective even in high-mass, high-density conditions ($M > \SI{e5}{\msol}, \Sigma > \SI{0.3}{\g\per\cm\squared}$), except for dust opacities corresponding to metallicities higher than found in Galactic GMCs (e.g. a few times Solar). This latter regime was studied by \citet{tsang2018} in a \SI{e7}{\msol} turbulent box, where radiation pressure only modestly reduced the star formation efficiency and rate, and did not disperse enough gas to stop star formation completely (helped by the porous gas distribution, in contrast with the \citet{fall2010} model). In short, the relatively low surface density and metallicity of our cloud (as well as the low luminosity) results in radiation pressure being unimportant as a source of feedback as it pertains to gas dispersal. For future models, we aim to include sink particle formation and accretion and therefore will be able to comment on how these feedback mechanisms affect star formation directly.

\subsection{Temperature and electron density}
\label{sec:m1e4-temperaturedensity}

\begin{figure}
    \centering
	\includegraphics[width=\columnwidth]{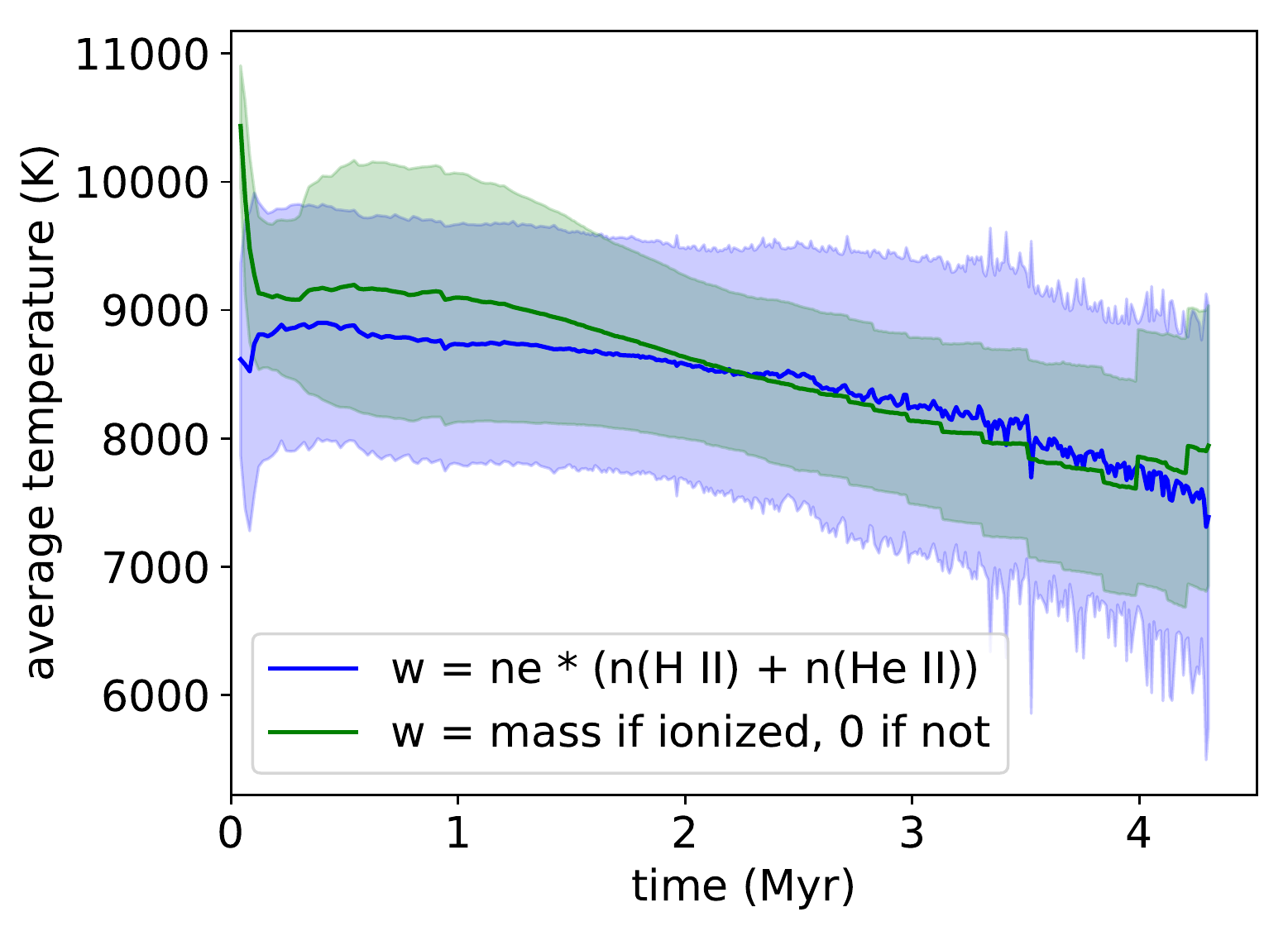}
    \caption{Volume-average ionized gas temperature as a function of time, weighted by $w = n_e (n_{\ion{H}{II}}+ n_{\ion{He}{II}}) \approx n_e^2$ (blue), mass $w=\rho \Delta V$ if the cell is more than 90 per cent ionized or $w=0$ if less (green). The filled edge shows the standard deviation. This is for the model with both feedback processes.}
    \label{fig:m1e4-avgtemp}
\end{figure}

\begin{figure}
    \centering
	\includegraphics[width=\columnwidth]{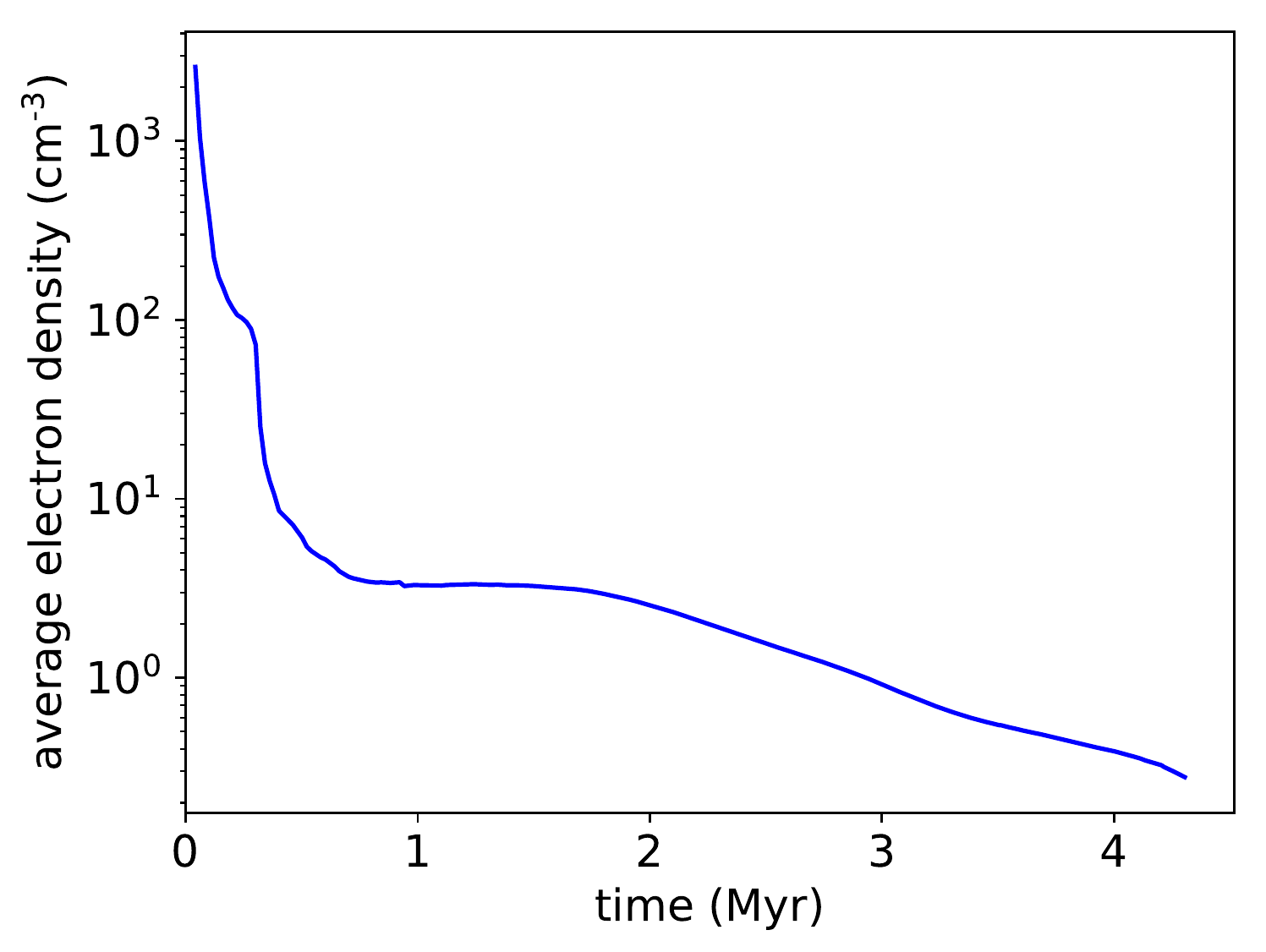}
    \caption{Volume-average electron density $n_e$ for ionized gas as a function of time in the model with both feedback processes.}
    \label{fig:m1e4-avgne}
\end{figure}

\Cref{fig:m1e4-avgtemp} and \Cref{fig:m1e4-avgne} show the average temperature and electron density, respectively, of the ionized gas, using two weighted averages: (a) $w = n_e (n_{\ion{H}{II}}+ n_{\ion{He}{II}}) \approx n_e^2$; (b) cell mass if hydrogen in the cell is more than 90 per cent ionized or $w=0$ if less. The results follow a similar pattern as the low-mass cloud, with both weighted-averages of the temperature decreasing down to \SI{8000}{K} by the end of the \SI{4}{\mega\yr}. The electron density goes down to \SI{0.3}{\per\cm\cubed}, again closely matching the low-mass cloud. This average density corresponds to a proton mass density of \SI{5e-25}{\g\per\cm\cubed}, which is five times more than the simulation floor density of \SI{e-25}{\g\per\cm\cubed}. The temperature and density (and thus thermal pressure) by the end of the simulation are as expected for the warm ionized medium \citep{wolfire1995}.


\section{FUV interstellar radiation field}
\label{sec:m1e4-fuvresults}

\begin{figure*}
    \centering
	\includegraphics[width=\textwidth]{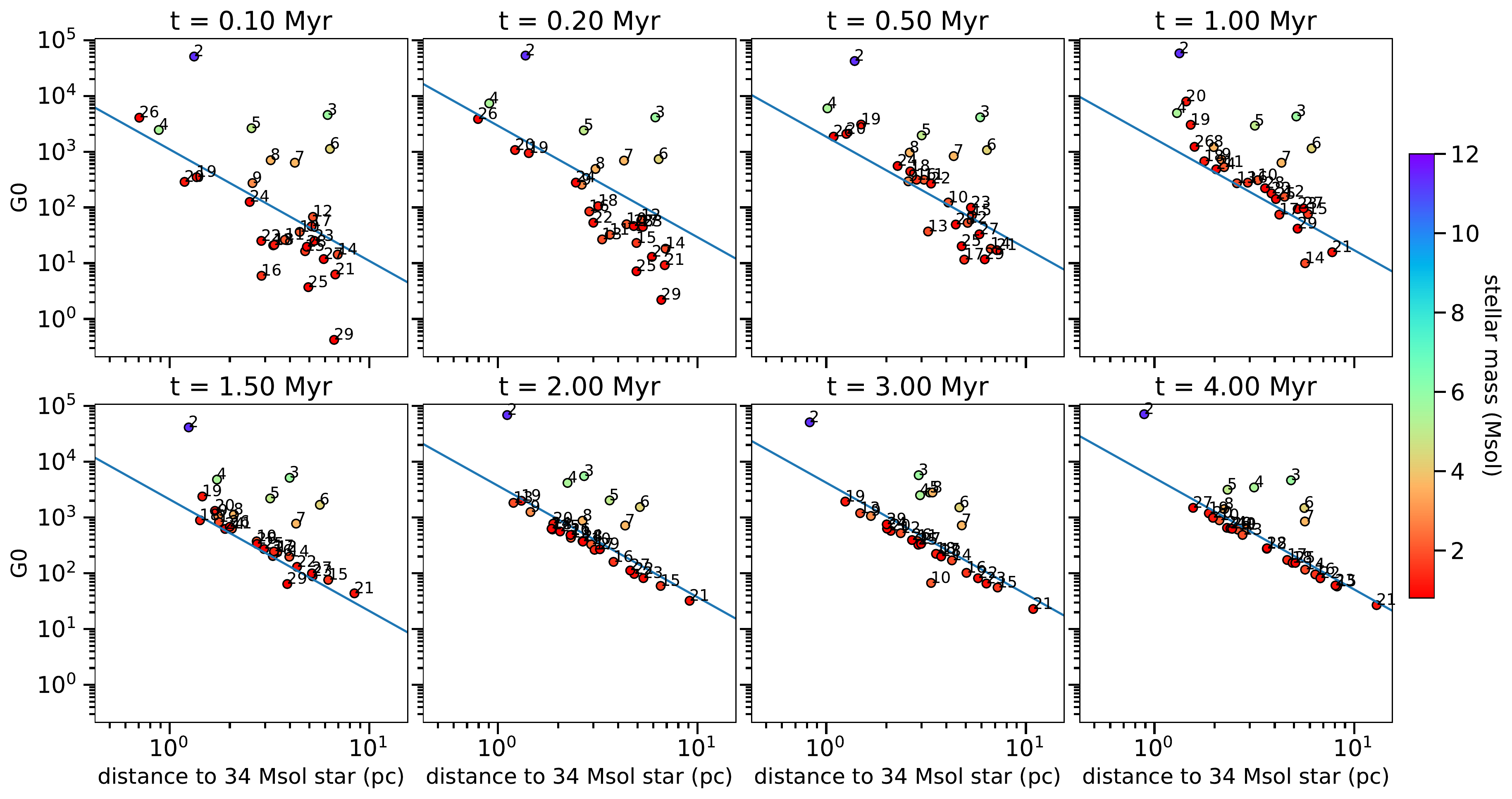}
	\caption{FUV interstellar radiation field $G_0$ (in units of the Habing field) as a function of distance from the most massive star, at different snapshots in time after feedback is initiated. Labels are shown for each star, corresponding to the labels in \cref{fig:m1e4-fuv}, with `2' being the second most massive star, and `29' the least massive. Colours show the stellar mass. The line shows the $r^{-2}$ dilution of $G_0$ from the cell with the most massive star.}
    \label{fig:distancevariation}
\end{figure*}

\Cref{fig:distancevariation} shows $G_0$ as a function of distance from the most massive star, at eight snapshots in time. Stars are labelled by a number going from 1 to 29 (where 1 is the most massive star, and 29 is the least massive). The corresponding (initial) stellar masses are labelled in \cref{fig:m1e4-fuv}. The straight line in \cref{fig:distancevariation} shows the purely geometric dilution of $G_0$ from the cell containing the massive star. There is less scatter after the first Myr, as the massive star disperses gas from the cluster, removing sources of FUV opacity, resulting in most of the stars falling along the straight line. There are some exceptions -- stars 2 to 8 (B stars from \SI{11}{\msol} to \SI{3.6}{\msol}) produce their own modest FUV luminosity which raises them above the massive star dilution line. In particular, star 2 has its own radius of influence ($\sim \SI{0.5}{\pc}$), in which some of the lower-mass stars (e.g. stars 19, 20, and 26) receive higher fluxes. Beyond this, star 1 dominates. At $t=\SI{3}{\mega\yr}$, star 10 is a factor of a few below the dilution line, due to a higher optical depth between it and star 1; this effect is more clearly seen in the plots of $G_0$ vs time, as discussed below.

The time-variation of the flux is plotted for every star in \cref{fig:m1e4-fuv}, alongside the integrated optical depth to the \SI{34}{\msol} star, which dominates the emission. The left axis shows $G_0$, where the solid line plots the spatial average within a radius of 2.5 cells as it varies with time; the dashed line shows the time-average value (with filled boundaries for the standard deviation). The right axis shows  $\exp (-\tau_\textrm{FUV})$ integrated along a ray between star 1 and the star of interest, where $\tau_\textrm{FUV}$ is given in \cref{eq:taufuv}. This is the path between the central cell of each sphere, so does not take into account any spatial averaging, but is a measure of the density moving into the ray directly. 

The results can be roughly split into three categories: stars which start off shielded but then become exposed with $G_0$ rapidly jumping up; stars which experience shielding later on such that $G_0$ either dips temporarily, or remains constant despite getting closer to the massive star; stars where the separation is the dominant factor in determining the variation in  $G_0$. These categories are not exclusive. In the following sections, we describe the former two categories, 

\begin{figure*}
	\includegraphics[width=\textwidth]{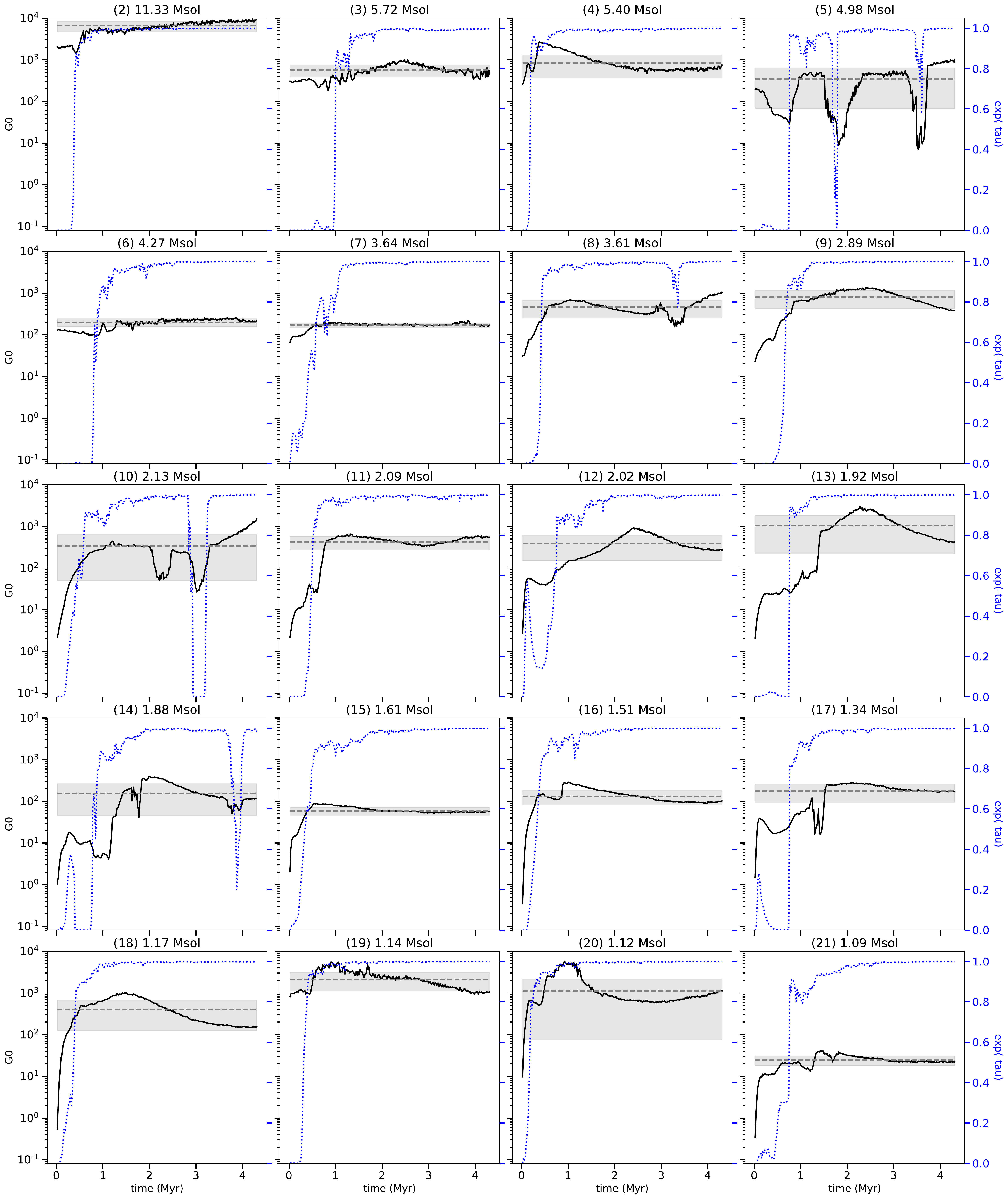}
    \caption{FUV interstellar radiation field $G_0$ in units of the Habing field for all non-massive stars (solid line, left axis). The dashed line shows the time-averaged $G_0$ with filled boundaries for the standard deviation. The dotted line (right axis) shows $\exp (-\tau_\textrm{FUV})$, where $\tau_\textrm{FUV}$ is the dust optical depth in the FUV integrated between the two stars. $G_0$ is extincted by moving clumps between the two moving stars. This shows the model with both feedback processes. Continues on next page.}
    \label{fig:m1e4-fuv}
\end{figure*}
\begin{figure*}
	\includegraphics[width=\textwidth]{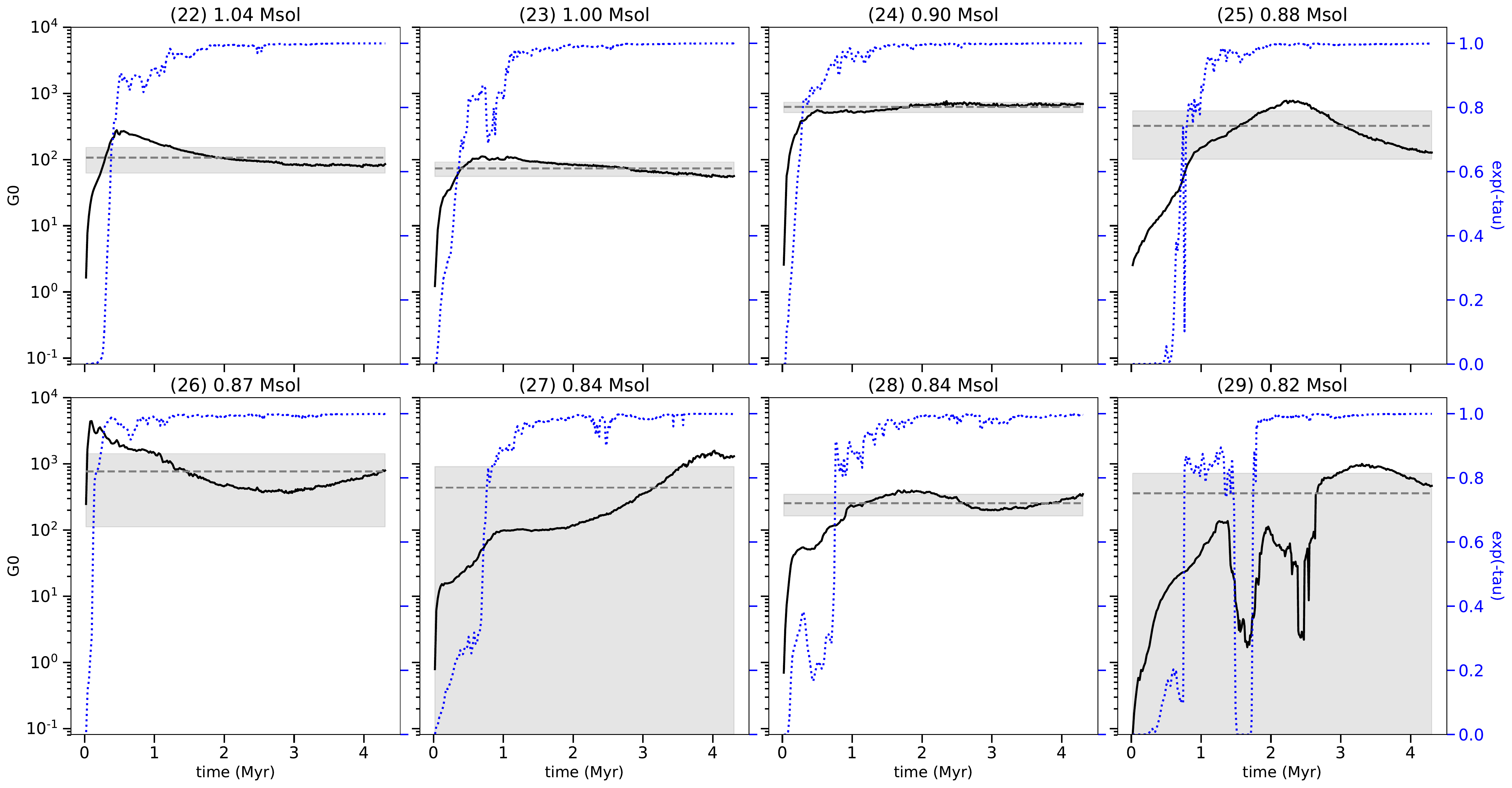}
    \contcaption{}
    \label{fig:m1e4-fuv-2}
\end{figure*}

\subsection{Sudden illumination}
\textbf{Star 7} begins \SI{4.25}{pc} away from the massive star (star 1), then moves to \SI{6}{pc} over \SI{4}{\mega\yr}. However, its value of $G_0$ is lowest at the beginning, despite this being its closest approach; it takes \SI{0.75}{\mega\yr} to rise by a factor of $\sim 3$ to $G_0=200$. This occurs as the massive star disperses material around itself, allowing its radiation field to reach more stars in the cluster with less intervening opacity. The column density between the two stars reduces around the \SI{4}{\mega\yr} mark, which results in a corresponding increase in $G_0$, despite the stars getting further away.

A similar evolution occurs for other stars, such as \textbf{star 4} and \textbf{star 15}, where the initial exposure takes \SI{0.5}{\mega\yr}. The latter's position is in one of the cavities blown out by the expanding \HII{} region (the bottom right quadrant of the plots shown in \cref{fig:m1e4-columndensity}), which decreases in column density. This means it does not take as long for $G_0$ to rise compared to star 7. It remains in the cavity with little intervening material, so there are no rapid rises or dips after the maximum value of $G_0 = 90$ is reached. \textbf{Stars 22} and \textbf{23} follow a similar pattern, with the latter positioned in the upper right quadrant cavity.

$G_0$ for \textbf{star 16} rises by an order of magnitude to 150 in the first \SI{0.45}{\mega\yr}, before slowly decreasing
down to 110 at $t=\SI{0.8}{\mega\yr}$, then jumping up to 280 in just \SI{0.12}{\mega\yr}. This is despite the star starting off at a distance of \SI{3}{\pc} and only getting further away. The delayed exposure is due to its location near the dense core in which star 1 is embedded. As that core expands and gets eroded, star 16 receives more flux; but due to the inhomogeneous distribution of gas, combined with the motion of the stars themselves, the column density increases, causing the slightly lower flux. Once this is itself dispersed, the star is more fully exposed.

This is even more the case for \textbf{star 20}, which is initially only \SI{1.2}{\pc} away from star 1. $G_0$ increases by a factor of 6 in the first \si{\mega\yr}, reaching a maximum value of 5700. 

The evolution of \textbf{star 21} is somewhat more complex, as it is located further away -- starting at \SI{7}{\pc} and increasing to \SI{13}{\pc}. As the \HII{} region expands, it pushes material in the way, so the illumination of star 21 occurs in a series of three steps up to \SI{1.45}{\mega\yr}. 

\textbf{Star 29} remains embedded in a dense filament (and later globule), which shields it up until \SI{2.5}{\mega\yr}, when the gas is photoevaporated away and the star becomes exposed. $G_0$ rises from 2 to 920 from $t=2.45$ to \SI{3.15}{\mega\yr} -- an increase of almost three orders of magnitude in the space of \SI{0.7}{\mega\yr}. 

\subsection{Temporary occultation}
\label{sec:transit}
The wide separation between \textbf{stars 21} and 1 (7 to \SI{13}{\pc}) means there is a greater chance for material to intervene: at \SI{1.68}{\mega\yr}, this causes a dip by a factor of 2 from $G_0 =40$ to 20. This is short-lived, as it rises after another \SI{0.1}{\mega\yr}.

\begin{figure}
    \centering
	\includegraphics[width=\columnwidth]{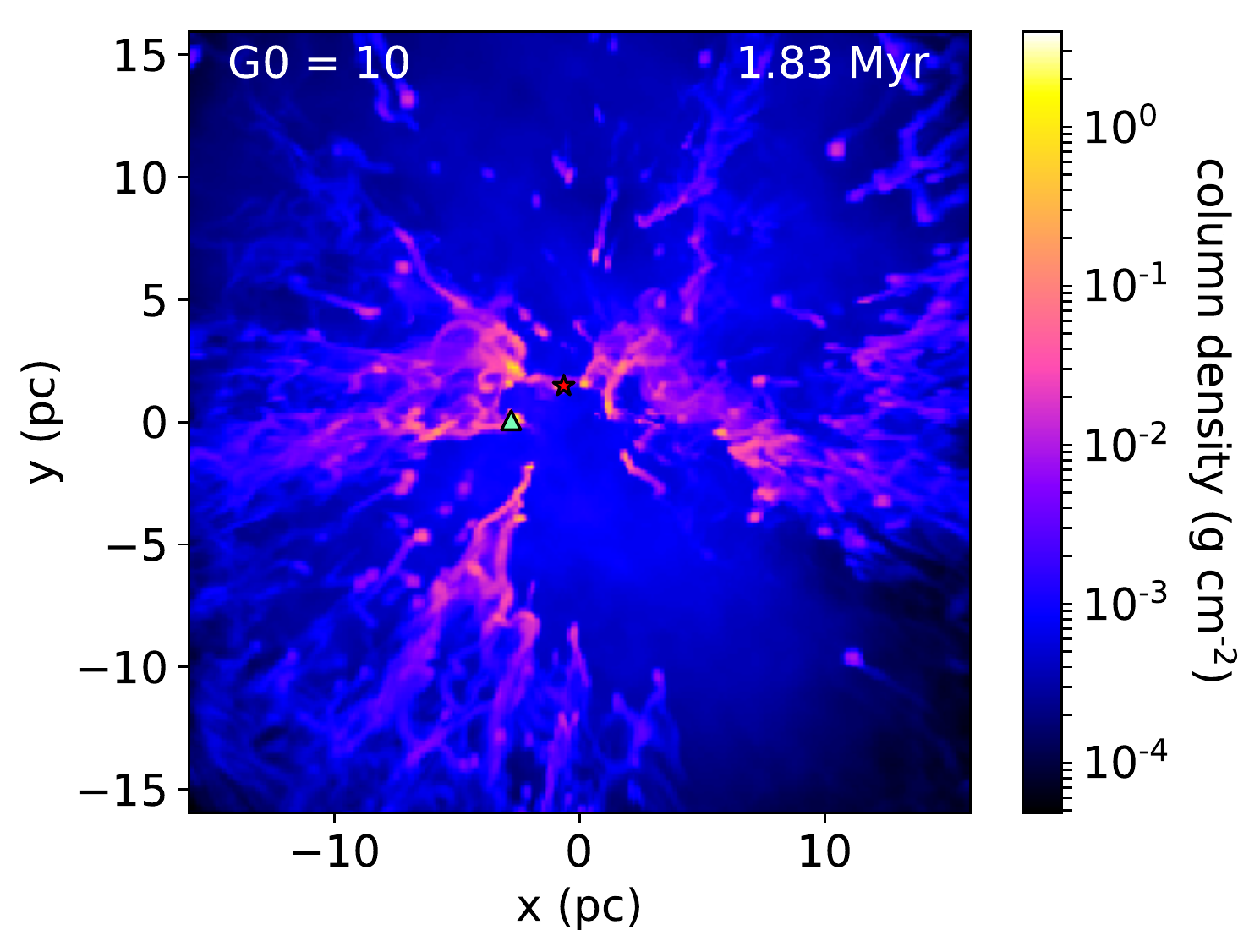}
    \caption[Column density snapshot with positions of massive star and star 5 with $G_0$]{Column density snapshot showing the positions of the \SI{34}{\msol} star (red star) and star 5, a \SI{4.98}{\msol} star (green triangle). The annotation displays $G_0$ at the location of star 5. Discussed in \cref{sec:transit}.}
    \label{fig:star5}
\end{figure}

\textbf{Star 5} is surrounded by a moving clump of gas which becomes aligned such that it blocks light from star 1. A column density snapshot is shown in \cref{fig:star5}. At \SI{1.5}{\mega\yr}, the flux decreases from 420 to 10, before being restored \SI{0.85}{\mega\yr} later. This happens again around \SI{3.5}{\mega\yr}, as $G_0$ drops from just over 400 to 7, a factor of 60 decrease. After \SI{0.2}{\mega\yr}, it then rapidly jumps up to 750 as the gas is blown away. This, combined with the narrower separation between the two stars, results in a final $G_0$ of 1000. \textbf{Star 10} follows a similar pattern to star 5 as it located near the same clump of gas. The occultation at $t=\SI{3}{\mega\yr}$ can be observed in the corresponding snapshot of \cref{fig:distancevariation}, which shows $G_0$ as a function of distance.

\subsection{Implications}
\begin{figure}
    \centering
	\includegraphics[width=\columnwidth]{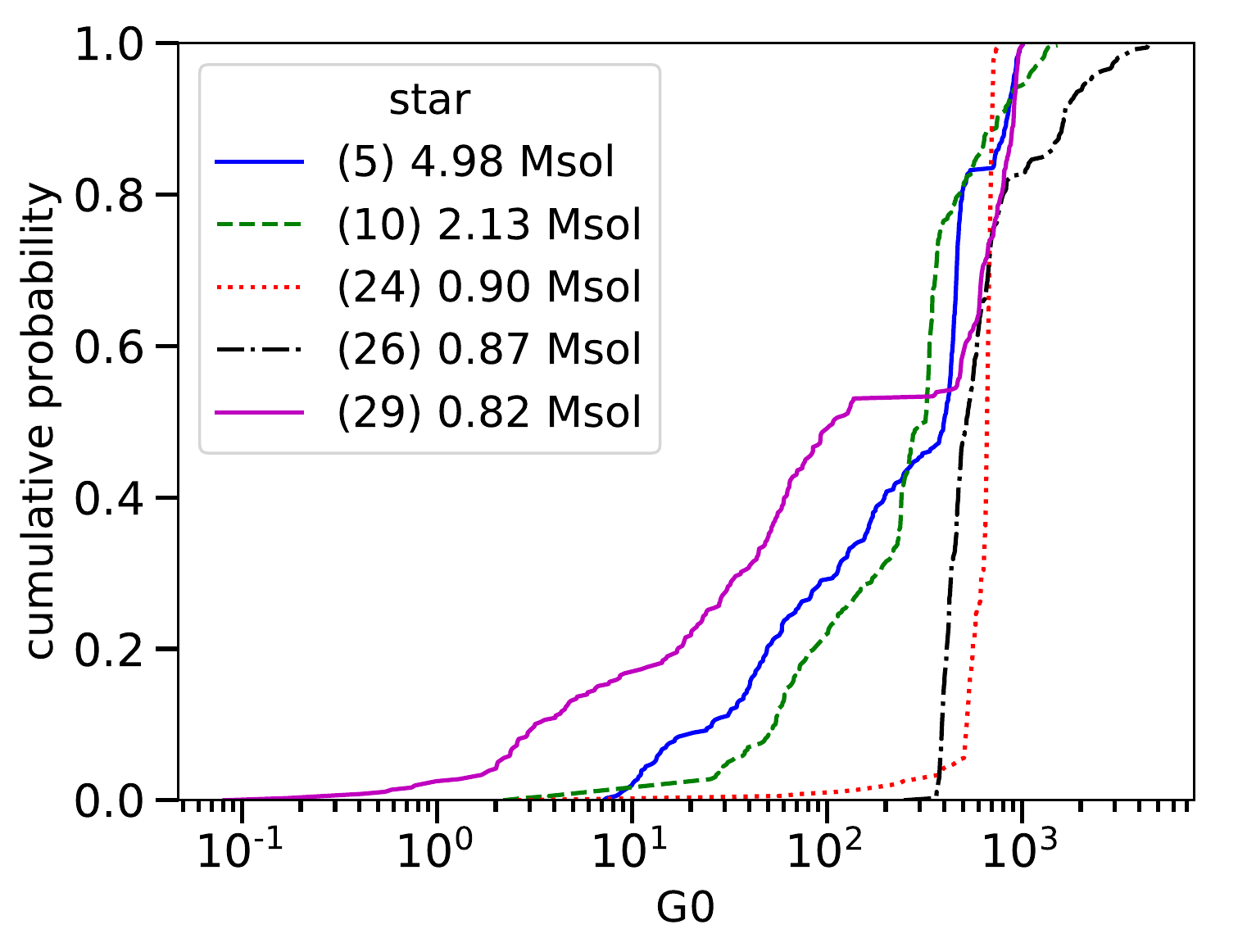}
    \caption{Cumulative probability distributions of $G_0$ for representative stars}
    \label{fig:cdf}
\end{figure}
The models presented here show how $G_0$ evolves in time as both gas and stars move around. During the first $\sim \SI{0.5}{\mega\yr}$ the most massive star is embedded within a core which is gradually eroded by the expanding \HII{} region -- shells expand outwards and have holes punched through them, allowing radiation to escape into the cluster medium. This means many stars are initially shielded from the massive star's radiation field, allowing them to evolve in relative isolation, but then become illuminated as material is dispersed. Some parcels of gas may move back in between, causing a drastic decrease in flux. 
In \cref{fig:cdf} we show cumulative probability distributions for some representative stars, showing how likely they are to have varying values of $G_0$. For example, star 24 (dotted red line) has a near-constant $G_0$ over its whole lifetime as it is not obstructed and stays at the same distance from the most massive star; star 26 (dot-dashed black line) is similar with the exception that the distance does vary ($\sim \SI{2}{\pc}$) -- this causes the 20 per cent variation at high $G_0$, Stars 5, 10, and 29 have larger ranges of $G_0$, with a preference to be found at the higher end, as their lows are caused by transient shielding.

In real star-forming regions, the interstellar radiation field can truncate and erode protoplanetary discs by photoevaporating them from the outside. Models by \citet{scally2001} and \citet{winter2018} show that disc destruction is dominated by photoevaporation over stellar encounters, making radiative feedback the primary external factor determining the fate of protoplanetary discs. Discs have lifetimes of the order of a few to \SI{10}{\mega\yr} \citep{haisch2001,mamajek2009,williams2011}, comparable to the main sequence lifetime for O stars. For example, \citet{adams2004} found that within time-scales of \SI{10}{\mega\yr}, discs around solar-mass stars can be severely truncated (down to radii below \SI{15}{AU}) with $G_0 = \SI{3e4}{}$, while stellar masses below \SI{0.5}{\msol} only require $G_0 = \SI{3e3}{}$. Such fields are present in the models presented here and in \citetalias{ali2018}. Even small values of $G_0$ can be significant, as shown by \citet{facchini2016} who calculated mass-loss rates of \SI{e-7}{\msol\per\yr} with $G_0 \sim 30$, provided grain growth is taken into account. Disc truncation will by necessity remove the possibility for planets to be formed outside that radius.

For the proplyds, the derived mass-loss rates of $\sim\SI{e-7}{}$ to \SI{e-5}{\msol\per\yr} imply short lifetimes of \SI{e4}{} to \SI{e5}{\yr} \citep{storzer1999,henney1999,brandner2000}. This gives rise to the so-called proplyd lifetime problem \citep[see e.g.][]{odell2001,scally2001,clarke2007}, as numerous proplyds are still observed in \HII{} regions. The most commonly proposed solution (assuming the mass measurements are accurate) is that the O stars are very young and have only just begun to shine on nearby objects, and the proplyds will be rapidly eroded away. However, it is unlikely that the observations of several \HII{} regions where proplyds are found are all capturing a special, short-lived time in their evolution and in the lifetime of the O star. A complementary explanation could lie in the stellar motions and transient shielding found in our model, both early on when the massive star is embedded, and later via occultation by clumps of gas and dust. This may prolong the proplyd lifetimes, as photoevaporation rates would not be constant.  A constant $G_0$, as used in the aforementioned models, implies a constant separation between disc and external source (as the geometric dilution does not change) as well as constant optical depth between them. These simulations show that the latter effect is particularly important in the first Myr (when the scatter is greatest in \cref{fig:distancevariation}, and the extinction factor $\ue^{-\tau}$ in \cref{fig:m1e4-fuv} is smallest), and afterwards the distance dependence becomes the dominant factor across the ensemble (when the scatter reduces, and $\ue^{-\tau} \sim 1$). Stars which do not experience any opacity effects can still have large deviations in flux arising from stellar motions (e.g. star 26). In summary, the cluster-gas interactions have a significant influence on the flux impinging on cluster members, and therefore our simulations can provide crucial input parameters for protoplanetary disc models which aim to calculate more realistic photoevaporative mass-loss rates and, when combined with viscous evolutionary models, realistic lifetimes.

\section{Summary and conclusions}
\label{sec:conclusions}
We have modelled a \SI{e4}{\msol} cloud with the same surface density and stellar mass function as \citet[][Paper I]{ali2018}, where a \SI{e3}{\msol} cloud was investigated. The same numerical methods are present in both models, allowing a comparison to be made between the two mass regimes. In summary:
\begin{enumerate}
    \item The higher mass cloud is somewhat more resistant to feedback, with 25 per cent of the initial mass remaining inside $(\SI{32.3}{\pc})^3$ after \SI{4.3}{\mega\yr}. In the lower-mass model, almost all material is removed from the $(\SI{15.5}{\pc})^3$ grid within \SI{1.6}{\mega\yr}. 
    \item Given that 75 per cent of the mass is removed, photoionization feedback is an effective mechanism for disrupting GMCs. Mass leaves the grid with a peak flux of \SI{4.7e-3}{\msol\per\yr}, comparable to the effects of supernovae in the model by \citet{rogers2013}.
    \item The total mass in Jeans-unstable cells approaches \SI{1500}{\msol}, representing a 15 per cent core formation efficiency or potential 5 per cent star formation efficiency. This estimate agrees with observational constraints.
    \item Radiation pressure has a negligible effect on bulk dispersal measures, although there is a minor difference in the total Jeans-unstable mass -- there is \SI{300}{\msol} more in the combined feedback model than the ionization-only model. This indicates radiation pressure may help to induce core formation.
    \item The time-variation of $G_0$ impinging on other cluster members is complex. It depends on the dispersal of the core in which the massive star is embedded (which takes just over \SI{0.5}{\mega\yr}, during which objects are shielded), and on occulations by dense ISM material. There can be drastic increases or decreases by orders of magnitude, lasting for durations of $\sim \SI{1}{\mega\yr}$. This may help to explain the proplyd lifetime problem by temporarily lowering photoevaporation rates.
    \item There is scatter in the relationship between incident $G_0$ and distance from the most massive star due to dust extinction. After the central cluster gas is dispersed, the scatter reduces and geometric dilution dominates for most stars. B stars produce a modest FUV field which keeps them above the diluted O-star flux, while occulted stars fall below this level.  
\end{enumerate}

The total run time of this model was \SI{4.3}{\mega\yr}, about a Myr before the \SI{34}{\msol} star will explode as a supernova (SN). A quarter of the gas still remains on the grid, and is unlikely to significantly disperse by the time the SN occurs, as the grid-boundary mass flux falls off over time. However, since this material is located in small, dense filaments and clumps, surrounded by large cavities of diffuse ionized gas, it is unlikely that the SN will transfer a significant proportion of its energy into the gas, whether it occurs now or in a Myr. We aim to investigate this in a future paper.

\section*{Acknowledgements}
We thank the referee for insightful and constructive comments. We thank Julian Pittard and Tim Naylor for comments which improved an early version of this paper. We also thank Tom Haworth and Matthew Bate for useful discussions.
AA is funded by the University of Exeter and a Science and Technology Facilities Council (STFC) studentship. TJH is funded by STFC Consolidated Grant ST/M00127X/1. 
The calculations for this paper were performed on
DiRAC Complexity and DiRAC Data Intensive (DIaL), both at the University of Leicester, and the DiRAC@Durham facility managed by the Institute for Computational Cosmology. These form part of the STFC DiRAC HPC Facility (www.dirac.ac.uk). 
Complexity was funded by BIS National E-Infrastructure capital grant ST/K000373/1 and  STFC DiRAC Operations grant ST/K0003259/1. 
DIaL was funded by BEIS capital funding via STFC capital grants ST/K000373/1 and ST/R002363/1 and STFC DiRAC Operations grant ST/R001014/1.
DiRAC@Durham was funded by BEIS capital funding via STFC capital grants ST/P002293/1, ST/R002371/1 and ST/S002502/1, Durham University and STFC operations grant ST/R000832/1. DiRAC is part of the National e-Infrastructure.

The research data supporting this publication are openly available from the University of Exeter's institutional repository at \href{https://ore.exeter.ac.uk/repository/}{https://ore.exeter.ac.uk/repository/}.




\bibliographystyle{mnras}
\bibliography{refs}

\begin{thebibliography}{}
\makeatletter
\relax
\def\mn@urlcharsother{\let\do\@makeother \do\$\do\&\do\#\do\^\do\_\do\%\do\~}
\def\mn@doi{\begingroup\mn@urlcharsother \@ifnextchar [ {\mn@doi@}
  {\mn@doi@[]}}
\def\mn@doi@[#1]#2{\def\@tempa{#1}\ifx\@tempa\@empty \href
  {http://dx.doi.org/#2} {doi:#2}\else \href {http://dx.doi.org/#2} {#1}\fi
  \endgroup}
\def\mn@eprint#1#2{\mn@eprint@#1:#2::\@nil}
\def\mn@eprint@arXiv#1{\href {http://arxiv.org/abs/#1} {{\tt arXiv:#1}}}
\def\mn@eprint@dblp#1{\href {http://dblp.uni-trier.de/rec/bibtex/#1.xml}
  {dblp:#1}}
\def\mn@eprint@#1:#2:#3:#4\@nil{\def\@tempa {#1}\def\@tempb {#2}\def\@tempc
  {#3}\ifx \@tempc \@empty \let \@tempc \@tempb \let \@tempb \@tempa \fi \ifx
  \@tempb \@empty \def\@tempb {arXiv}\fi \@ifundefined
  {mn@eprint@\@tempb}{\@tempb:\@tempc}{\expandafter \expandafter \csname
  mn@eprint@\@tempb\endcsname \expandafter{\@tempc}}}

\bibitem[\protect\citeauthoryear{{Adams}, {Hollenbach}, {Laughlin}  \&
  {Gorti}}{{Adams} et~al.}{2004}]{adams2004}
{Adams} F.~C.,  {Hollenbach} D.,  {Laughlin} G.,   {Gorti} U.,  2004, \mn@doi
  [\apj] {10.1086/421989}, \href
  {http://adsabs.harvard.edu/abs/2004ApJ...611..360A} {611, 360}

\bibitem[\protect\citeauthoryear{{Ali}, {Harries}  \& {Douglas}}{{Ali}
  et~al.}{2018}]{ali2018}
{Ali} A.,  {Harries} T.~J.,   {Douglas} T.~A.,  2018, \mn@doi [\mnras]
  {10.1093/mnras/sty1001}, \href
  {http://adsabs.harvard.edu/abs/2018MNRAS.477.5422A} {477, 5422}

\bibitem[\protect\citeauthoryear{{Alves}, {Lombardi}  \& {Lada}}{{Alves}
  et~al.}{2007}]{alves2007}
{Alves} J.,  {Lombardi} M.,   {Lada} C.~J.,  2007, \mn@doi [\aap]
  {10.1051/0004-6361:20066389}, \href
  {http://adsabs.harvard.edu/abs/2007A%26A...462L..17A} {462, L17}

\bibitem[\protect\citeauthoryear{{Anderson}, {Adams}  \& {Calvet}}{{Anderson}
  et~al.}{2013}]{anderson2013}
{Anderson} K.~R.,  {Adams} F.~C.,   {Calvet} N.,  2013, \mn@doi [\apj]
  {10.1088/0004-637X/774/1/9}, \href
  {http://adsabs.harvard.edu/abs/2013ApJ...774....9A} {774, 9}

\bibitem[\protect\citeauthoryear{{Ansdell}, {Williams}, {Manara}, {Miotello},
  {Facchini}, {van der Marel}, {Testi}  \& {van Dishoeck}}{{Ansdell}
  et~al.}{2017}]{ansdell2017}
{Ansdell} M.,  {Williams} J.~P.,  {Manara} C.~F.,  {Miotello} A.,  {Facchini}
  S.,  {van der Marel} N.,  {Testi} L.,   {van Dishoeck} E.~F.,  2017, \mn@doi
  [\aj] {10.3847/1538-3881/aa69c0}, \href
  {http://adsabs.harvard.edu/abs/2017AJ....153..240A} {153, 240}

\bibitem[\protect\citeauthoryear{{Bally}, {O'Dell}  \& {McCaughrean}}{{Bally}
  et~al.}{2000}]{bally2000}
{Bally} J.,  {O'Dell} C.~R.,   {McCaughrean} M.~J.,  2000, \mn@doi [\aj]
  {10.1086/301385}, \href {http://adsabs.harvard.edu/abs/2000AJ....119.2919B}
  {119, 2919}

\bibitem[\protect\citeauthoryear{{Bate}, {Bonnell}  \& {Bromm}}{{Bate}
  et~al.}{2002}]{bate2002}
{Bate} M.~R.,  {Bonnell} I.~A.,   {Bromm} V.,  2002, \mn@doi [\mnras]
  {10.1046/j.1365-8711.2002.05539.x}, \href
  {http://adsabs.harvard.edu/abs/2002MNRAS.332L..65B} {332, L65}

\bibitem[\protect\citeauthoryear{{Bodenheimer}, {Tenorio-Tagle}  \&
  {Yorke}}{{Bodenheimer} et~al.}{1979}]{bodenheimer1979}
{Bodenheimer} P.,  {Tenorio-Tagle} G.,   {Yorke} H.~W.,  1979, \mn@doi [\apj]
  {10.1086/157368}, \href {http://adsabs.harvard.edu/abs/1979ApJ...233...85B}
  {233, 85}

\bibitem[\protect\citeauthoryear{{Brandner} et~al.,}{{Brandner}
  et~al.}{2000}]{brandner2000}
{Brandner} W.,  et~al., 2000, \mn@doi [\aj] {10.1086/301192}, \href
  {http://adsabs.harvard.edu/abs/2000AJ....119..292B} {119, 292}

\bibitem[\protect\citeauthoryear{{Clarke}}{{Clarke}}{2007}]{clarke2007}
{Clarke} C.~J.,  2007, \mn@doi [\mnras] {10.1111/j.1365-2966.2007.11547.x},
  \href {http://adsabs.harvard.edu/abs/2007MNRAS.376.1350C} {376, 1350}

\bibitem[\protect\citeauthoryear{{Dale} \& {Bonnell}}{{Dale} \&
  {Bonnell}}{2011}]{dale2011}
{Dale} J.~E.,  {Bonnell} I.,  2011, \mn@doi [\mnras]
  {10.1111/j.1365-2966.2011.18392.x}, \href
  {http://adsabs.harvard.edu/abs/2011MNRAS.414..321D} {414, 321}

\bibitem[\protect\citeauthoryear{{Dale}, {Bonnell}, {Clarke}  \& {Bate}}{{Dale}
  et~al.}{2005}]{dale2005}
{Dale} J.~E.,  {Bonnell} I.~A.,  {Clarke} C.~J.,   {Bate} M.~R.,  2005, \mn@doi
  [\mnras] {10.1111/j.1365-2966.2005.08806.x}, \href
  {http://adsabs.harvard.edu/abs/2005MNRAS.358..291D} {358, 291}

\bibitem[\protect\citeauthoryear{{Dale}, {Ercolano}  \& {Bonnell}}{{Dale}
  et~al.}{2012}]{dale2012}
{Dale} J.~E.,  {Ercolano} B.,   {Bonnell} I.~A.,  2012, \mn@doi [\mnras]
  {10.1111/j.1365-2966.2012.21205.x}, \href
  {http://adsabs.harvard.edu/abs/2012MNRAS.424..377D} {424, 377}

\bibitem[\protect\citeauthoryear{{Dale}, {Ercolano}  \& {Bonnell}}{{Dale}
  et~al.}{2013}]{dale2013a}
{Dale} J.~E.,  {Ercolano} B.,   {Bonnell} I.~A.,  2013, \mn@doi [\mnras]
  {10.1093/mnras/sts592}, \href
  {http://adsabs.harvard.edu/abs/2013MNRAS.430..234D} {430, 234}

\bibitem[\protect\citeauthoryear{{Draine}}{{Draine}}{1978}]{draine1978}
{Draine} B.~T.,  1978, \mn@doi [\apjs] {10.1086/190513}, \href
  {http://adsabs.harvard.edu/abs/1978ApJS...36..595D} {36, 595}

\bibitem[\protect\citeauthoryear{{Draine} \& {Lee}}{{Draine} \&
  {Lee}}{1984}]{draine1984}
{Draine} B.~T.,  {Lee} H.~M.,  1984, \mn@doi [\apj] {10.1086/162480}, \href
  {http://adsabs.harvard.edu/abs/1984ApJ...285...89D} {285, 89}

\bibitem[\protect\citeauthoryear{{Eisner} et~al.,}{{Eisner}
  et~al.}{2018}]{eisner2018}
{Eisner} J.~A.,  et~al., 2018, \mn@doi [\apj] {10.3847/1538-4357/aac3e2}, \href
  {http://adsabs.harvard.edu/abs/2018ApJ...860...77E} {860, 77}

\bibitem[\protect\citeauthoryear{{Ercolano} \& {Gritschneder}}{{Ercolano} \&
  {Gritschneder}}{2011}]{ercolano2011}
{Ercolano} B.,  {Gritschneder} M.,  2011, \mn@doi [\mnras]
  {10.1111/j.1365-2966.2011.18144.x}, \href
  {http://adsabs.harvard.edu/abs/2011MNRAS.413..401E} {413, 401}

\bibitem[\protect\citeauthoryear{{Facchini}, {Clarke}  \& {Bisbas}}{{Facchini}
  et~al.}{2016}]{facchini2016}
{Facchini} S.,  {Clarke} C.~J.,   {Bisbas} T.~G.,  2016, \mn@doi [\mnras]
  {10.1093/mnras/stw240}, \href
  {http://adsabs.harvard.edu/abs/2016MNRAS.457.3593F} {457, 3593}

\bibitem[\protect\citeauthoryear{{Fall}, {Krumholz}  \& {Matzner}}{{Fall}
  et~al.}{2010}]{fall2010}
{Fall} S.~M.,  {Krumholz} M.~R.,   {Matzner} C.~D.,  2010, \mn@doi [\apjl]
  {10.1088/2041-8205/710/2/L142}, \href
  {http://adsabs.harvard.edu/abs/2010ApJ...710L.142F} {710, L142}

\bibitem[\protect\citeauthoryear{{Franco}, {Tenorio-Tagle}  \&
  {Bodenheimer}}{{Franco} et~al.}{1990}]{franco1990}
{Franco} J.,  {Tenorio-Tagle} G.,   {Bodenheimer} P.,  1990, \mn@doi [\apj]
  {10.1086/168300}, \href {http://adsabs.harvard.edu/abs/1990ApJ...349..126F}
  {349, 126}

\bibitem[\protect\citeauthoryear{{Geen}, {Hennebelle}, {Tremblin}  \&
  {Rosdahl}}{{Geen} et~al.}{2016}]{geen2016}
{Geen} S.,  {Hennebelle} P.,  {Tremblin} P.,   {Rosdahl} J.,  2016, \mn@doi
  [\mnras] {10.1093/mnras/stw2235}, \href
  {http://adsabs.harvard.edu/abs/2016MNRAS.463.3129G} {463, 3129}

\bibitem[\protect\citeauthoryear{{Geen}, {Soler}  \& {Hennebelle}}{{Geen}
  et~al.}{2017}]{geen2017}
{Geen} S.,  {Soler} J.~D.,   {Hennebelle} P.,  2017, \mn@doi [\mnras]
  {10.1093/mnras/stx1765}, \href
  {http://adsabs.harvard.edu/abs/2017MNRAS.471.4844G} {471, 4844}

\bibitem[\protect\citeauthoryear{{Geen}, {Watson}, {Rosdahl}, {Bieri},
  {Klessen}  \& {Hennebelle}}{{Geen} et~al.}{2018}]{geen2018}
{Geen} S.,  {Watson} S.~K.,  {Rosdahl} J.,  {Bieri} R.,  {Klessen} R.~S.,
  {Hennebelle} P.,  2018, \mn@doi [\mnras] {10.1093/mnras/sty2439}, \href
  {http://adsabs.harvard.edu/abs/2018MNRAS.481.2548G} {481, 2548}

\bibitem[\protect\citeauthoryear{{Guarcello}, {Prisinzano}, {Micela},
  {Damiani}, {Peres}  \& {Sciortino}}{{Guarcello} et~al.}{2007}]{guarcello2007}
{Guarcello} M.~G.,  {Prisinzano} L.,  {Micela} G.,  {Damiani} F.,  {Peres} G.,
   {Sciortino} S.,  2007, \mn@doi [\aap] {10.1051/0004-6361:20066124}, \href
  {http://adsabs.harvard.edu/abs/2007A%26A...462..245G} {462, 245}

\bibitem[\protect\citeauthoryear{{Guarcello}, {Micela}, {Peres}, {Prisinzano}
  \& {Sciortino}}{{Guarcello} et~al.}{2010}]{guarcello2010}
{Guarcello} M.~G.,  {Micela} G.,  {Peres} G.,  {Prisinzano} L.,   {Sciortino}
  S.,  2010, \mn@doi [\aap] {10.1051/0004-6361/201014351}, \href
  {http://adsabs.harvard.edu/abs/2010A%26A...521A..61G} {521, A61}

\bibitem[\protect\citeauthoryear{{Guarcello} et~al.,}{{Guarcello}
  et~al.}{2016}]{guarcello2016}
{Guarcello} M.~G.,  et~al., 2016, preprint, \href
  {http://adsabs.harvard.edu/abs/2016arXiv160501773G} {} (\mn@eprint {}
  {1605.01773})

\bibitem[\protect\citeauthoryear{{Habing}}{{Habing}}{1968}]{habing1968}
{Habing} H.~J.,  1968, \bain, \href
  {http://adsabs.harvard.edu/abs/1968BAN....19..421H} {19, 421}

\bibitem[\protect\citeauthoryear{{Haisch}, {Lada}  \& {Lada}}{{Haisch}
  et~al.}{2001}]{haisch2001}
{Haisch} Jr. K.~E.,  {Lada} E.~A.,   {Lada} C.~J.,  2001, \mn@doi [\apjl]
  {10.1086/320685}, \href {http://adsabs.harvard.edu/abs/2001ApJ...553L.153H}
  {553, L153}

\bibitem[\protect\citeauthoryear{{Harries}}{{Harries}}{2015}]{harries2015}
{Harries} T.~J.,  2015, \mn@doi [\mnras] {10.1093/mnras/stv158}, \href
  {http://adsabs.harvard.edu/abs/2015MNRAS.448.3156H} {448, 3156}

\bibitem[\protect\citeauthoryear{{Harries}, {Haworth}, {Acreman}, {Ali}  \&
  {Douglas}}{{Harries} et~al.}{2019}]{harries2019}
{Harries} T.~J.,  {Haworth} T.~J.,  {Acreman} D.,  {Ali} A.,   {Douglas} T.,
  2019, \mn@doi [Astronomy and Computing] {10.1016/j.ascom.2019.03.002}, \href
  {http://adsabs.harvard.edu/abs/2019A%26C....27...63H} {27, 63}

\bibitem[\protect\citeauthoryear{{Haworth} \& {Harries}}{{Haworth} \&
  {Harries}}{2012}]{haworth2012}
{Haworth} T.~J.,  {Harries} T.~J.,  2012, \mn@doi [\mnras]
  {10.1111/j.1365-2966.2011.20062.x}, \href
  {http://adsabs.harvard.edu/abs/2012MNRAS.420..562H} {420, 562}

\bibitem[\protect\citeauthoryear{{Haworth}, {Harries}, {Acreman}  \&
  {Bisbas}}{{Haworth} et~al.}{2015}]{haworth2015}
{Haworth} T.~J.,  {Harries} T.~J.,  {Acreman} D.~M.,   {Bisbas} T.~G.,  2015,
  \mn@doi [\mnras] {10.1093/mnras/stv1814}, \href
  {http://adsabs.harvard.edu/abs/2015MNRAS.453.2277H} {453, 2277}

\bibitem[\protect\citeauthoryear{{Haworth}, {Facchini}, {Clarke}  \&
  {Cleeves}}{{Haworth} et~al.}{2017}]{haworth2017}
{Haworth} T.~J.,  {Facchini} S.,  {Clarke} C.~J.,   {Cleeves} L.~I.,  2017,
  \mn@doi [\mnras] {10.1093/mnrasl/slx037}, \href
  {http://adsabs.harvard.edu/abs/2017MNRAS.468L.108H} {468, L108}

\bibitem[\protect\citeauthoryear{{Haworth}, {Clarke}, {Rahman}, {Winter}  \&
  {Facchini}}{{Haworth} et~al.}{2018}]{haworth2018c}
{Haworth} T.~J.,  {Clarke} C.~J.,  {Rahman} W.,  {Winter} A.~J.,   {Facchini}
  S.,  2018, \mn@doi [\mnras] {10.1093/mnras/sty2323}, \href
  {http://adsabs.harvard.edu/abs/2018MNRAS.tmp.2197H} {481, 452}

\bibitem[\protect\citeauthoryear{{Henney} \& {O'Dell}}{{Henney} \&
  {O'Dell}}{1999}]{henney1999}
{Henney} W.~J.,  {O'Dell} C.~R.,  1999, \mn@doi [\aj] {10.1086/301087}, \href
  {http://adsabs.harvard.edu/abs/1999AJ....118.2350H} {118, 2350}

\bibitem[\protect\citeauthoryear{{Heyer}, {Krawczyk}, {Duval}  \&
  {Jackson}}{{Heyer} et~al.}{2009}]{heyer2009}
{Heyer} M.,  {Krawczyk} C.,  {Duval} J.,   {Jackson} J.~M.,  2009, \mn@doi
  [\apj] {10.1088/0004-637X/699/2/1092}, \href
  {http://adsabs.harvard.edu/abs/2009ApJ...699.1092H} {699, 1092}

\bibitem[\protect\citeauthoryear{{Hollenbach} \& {McKee}}{{Hollenbach} \&
  {McKee}}{1979}]{hollenbach1979}
{Hollenbach} D.,  {McKee} C.~F.,  1979, \mn@doi [\apjs] {10.1086/190631}, \href
  {http://adsabs.harvard.edu/abs/1979ApJS...41..555H} {41, 555}

\bibitem[\protect\citeauthoryear{{Hollenbach} \& {Tielens}}{{Hollenbach} \&
  {Tielens}}{1999}]{hollenbach1999}
{Hollenbach} D.~J.,  {Tielens} A.~G.~G.~M.,  1999, \mn@doi [Reviews of Modern
  Physics] {10.1103/RevModPhys.71.173}, \href
  {http://adsabs.harvard.edu/abs/1999RvMP...71..173H} {71, 173}

\bibitem[\protect\citeauthoryear{{Howard}, {Pudritz}  \& {Harris}}{{Howard}
  et~al.}{2017}]{howard2017a}
{Howard} C.~S.,  {Pudritz} R.~E.,   {Harris} W.~E.,  2017, \mn@doi [\mnras]
  {10.1093/mnras/stx1363}, \href
  {http://adsabs.harvard.edu/abs/2017MNRAS.470.3346H} {470, 3346}

\bibitem[\protect\citeauthoryear{{Johnstone}, {Hollenbach}  \&
  {Bally}}{{Johnstone} et~al.}{1998}]{johnstone1998}
{Johnstone} D.,  {Hollenbach} D.,   {Bally} J.,  1998, \mn@doi [\apj]
  {10.1086/305658}, \href {http://adsabs.harvard.edu/abs/1998ApJ...499..758J}
  {499, 758}

\bibitem[\protect\citeauthoryear{{Kim}, {Clarke}, {Fang}  \& {Facchini}}{{Kim}
  et~al.}{2016}]{kim2016a}
{Kim} J.~S.,  {Clarke} C.~J.,  {Fang} M.,   {Facchini} S.,  2016, \mn@doi
  [\apjl] {10.3847/2041-8205/826/1/L15}, \href
  {http://adsabs.harvard.edu/abs/2016ApJ...826L..15K} {826, L15}

\bibitem[\protect\citeauthoryear{{Kim}, {Kim}  \& {Ostriker}}{{Kim}
  et~al.}{2018}]{kim2018a}
{Kim} J.-G.,  {Kim} W.-T.,   {Ostriker} E.~C.,  2018, \apj, \href
  {http://adsabs.harvard.edu/abs/2018ApJ...859...68K} {859, 68}

\bibitem[\protect\citeauthoryear{{Kurucz}}{{Kurucz}}{1991}]{kurucz1991}
{Kurucz} R.~L.,  1991, in {Crivellari} L.,  {Hubeny} I.,   {Hummer} D.~G.,
  eds,  NATO Advanced Science Institutes (ASI) Series C Vol. 341, NATO Advanced
  Science Institutes (ASI) Series C. p.~441

\bibitem[\protect\citeauthoryear{{Lada} \& {Lada}}{{Lada} \&
  {Lada}}{2003}]{lada2003}
{Lada} C.~J.,  {Lada} E.~A.,  2003, \mn@doi [\araa]
  {10.1146/annurev.astro.41.011802.094844}, \href
  {http://adsabs.harvard.edu/abs/2003ARA%26A..41...57L} {41, 57}

\bibitem[\protect\citeauthoryear{{Lanz} \& {Hubeny}}{{Lanz} \&
  {Hubeny}}{2003}]{lanz2003}
{Lanz} T.,  {Hubeny} I.,  2003, \mn@doi [\apjs] {10.1086/374373}, \href
  {http://adsabs.harvard.edu/abs/2003ApJS..146..417L} {146, 417}

\bibitem[\protect\citeauthoryear{{Lucy}}{{Lucy}}{1999}]{lucy1999}
{Lucy} L.~B.,  1999, \aap, \href
  {http://adsabs.harvard.edu/abs/1999A%26A...344..282L} {344, 282}

\bibitem[\protect\citeauthoryear{{Mamajek}}{{Mamajek}}{2009}]{mamajek2009}
{Mamajek} E.~E.,  2009, in {Usuda} T.,  {Tamura} M.,   {Ishii} M.,  eds,
  American Institute of Physics Conference Series Vol. 1158, American Institute
  of Physics Conference Series. pp 3--10 (\mn@eprint {} {0906.5011}),
  \mn@doi{10.1063/1.3215910}

\bibitem[\protect\citeauthoryear{{Mathis}, {Rumpl}  \& {Nordsieck}}{{Mathis}
  et~al.}{1977}]{mathis1977}
{Mathis} J.~S.,  {Rumpl} W.,   {Nordsieck} K.~H.,  1977, \mn@doi [\apj]
  {10.1086/155591}, \href {http://adsabs.harvard.edu/abs/1977ApJ...217..425M}
  {217, 425}

\bibitem[\protect\citeauthoryear{{Matzner}}{{Matzner}}{2002}]{matzner2002}
{Matzner} C.~D.,  2002, \mn@doi [\apj] {10.1086/338030}, \href
  {http://adsabs.harvard.edu/abs/2002ApJ...566..302M} {566, 302}

\bibitem[\protect\citeauthoryear{{O'dell}}{{O'dell}}{1998}]{odell1998}
{O'dell} C.~R.,  1998, \mn@doi [\aj] {10.1086/300178}, \href
  {http://adsabs.harvard.edu/abs/1998AJ....115..263O} {115, 263}

\bibitem[\protect\citeauthoryear{{O'dell}}{{O'dell}}{2001}]{odell2001}
{O'dell} C.~R.,  2001, \mn@doi [\araa] {10.1146/annurev.astro.39.1.99}, \href
  {http://adsabs.harvard.edu/abs/2001ARA%26A..39...99O} {39, 99}

\bibitem[\protect\citeauthoryear{{O'dell}, {Wen}  \& {Hu}}{{O'dell}
  et~al.}{1993}]{odell1993}
{O'dell} C.~R.,  {Wen} Z.,   {Hu} X.,  1993, \mn@doi [\apj] {10.1086/172786},
  \href {http://adsabs.harvard.edu/abs/1993ApJ...410..696O} {410, 696}

\bibitem[\protect\citeauthoryear{{Rogers} \& {Pittard}}{{Rogers} \&
  {Pittard}}{2013}]{rogers2013}
{Rogers} H.,  {Pittard} J.~M.,  2013, \mn@doi [\mnras] {10.1093/mnras/stt255},
  \href {http://adsabs.harvard.edu/abs/2013MNRAS.431.1337R} {431, 1337}

\bibitem[\protect\citeauthoryear{{Scally} \& {Clarke}}{{Scally} \&
  {Clarke}}{2001}]{scally2001}
{Scally} A.,  {Clarke} C.,  2001, \mn@doi [\mnras]
  {10.1046/j.1365-8711.2001.04274.x}, \href
  {http://adsabs.harvard.edu/abs/2001MNRAS.325..449S} {325, 449}

\bibitem[\protect\citeauthoryear{{Schaller}, {Schaerer}, {Meynet}  \&
  {Maeder}}{{Schaller} et~al.}{1992}]{schaller1992}
{Schaller} G.,  {Schaerer} D.,  {Meynet} G.,   {Maeder} A.,  1992, \aaps, \href
  {http://adsabs.harvard.edu/abs/1992A%26AS...96..269S} {96, 269}

\bibitem[\protect\citeauthoryear{{Skinner} \& {Ostriker}}{{Skinner} \&
  {Ostriker}}{2015}]{skinner2015}
{Skinner} M.~A.,  {Ostriker} E.~C.,  2015, \mn@doi [\apj]
  {10.1088/0004-637X/809/2/187}, \href
  {http://adsabs.harvard.edu/abs/2015ApJ...809..187S} {809, 187}

\bibitem[\protect\citeauthoryear{{Smith}, {Bally}  \& {Morse}}{{Smith}
  et~al.}{2003}]{smith2003}
{Smith} N.,  {Bally} J.,   {Morse} J.~A.,  2003, \mn@doi [\apjl]
  {10.1086/375312}, \href {http://adsabs.harvard.edu/abs/2003ApJ...587L.105S}
  {587, L105}

\bibitem[\protect\citeauthoryear{{Solomon}, {Rivolo}, {Barrett}  \&
  {Yahil}}{{Solomon} et~al.}{1987}]{solomon1987}
{Solomon} P.~M.,  {Rivolo} A.~R.,  {Barrett} J.,   {Yahil} A.,  1987, \mn@doi
  [\apj] {10.1086/165493}, \href
  {http://adsabs.harvard.edu/abs/1987ApJ...319..730S} {319, 730}

\bibitem[\protect\citeauthoryear{{Stecher} \& {Williams}}{{Stecher} \&
  {Williams}}{1967}]{stecher1967}
{Stecher} T.~P.,  {Williams} D.~A.,  1967, \mn@doi [\apjl] {10.1086/180047},
  \href {http://adsabs.harvard.edu/abs/1967ApJ...149L..29S} {149, L29}

\bibitem[\protect\citeauthoryear{{St\"orzer} \& {Hollenbach}}{{St\"orzer} \&
  {Hollenbach}}{1999}]{storzer1999}
{St\"orzer} H.,  {Hollenbach} D.,  1999, \mn@doi [\apj] {10.1086/307055}, \href
  {http://adsabs.harvard.edu/abs/1999ApJ...515..669S} {515, 669}

\bibitem[\protect\citeauthoryear{{Tenorio-Tagle}}{{Tenorio-Tagle}}{1979}]{tenorio-tagle1979}
{Tenorio-Tagle} G.,  1979, \aap, \href
  {http://adsabs.harvard.edu/abs/1979A%26A....71...59T} {71, 59}

\bibitem[\protect\citeauthoryear{{Truelove}, {Klein}, {McKee}, {Holliman},
  {Howell}  \& {Greenough}}{{Truelove} et~al.}{1997}]{truelove1997}
{Truelove} J.~K.,  {Klein} R.~I.,  {McKee} C.~F.,  {Holliman} II J.~H.,
  {Howell} L.~H.,   {Greenough} J.~A.,  1997, \mn@doi [\apjl] {10.1086/310975},
  \href {http://adsabs.harvard.edu/abs/1997ApJ...489L.179T} {489, L179}

\bibitem[\protect\citeauthoryear{{Tsang} \& {Milosavljevi\'c}}{{Tsang} \&
  {Milosavljevi\'c}}{2018}]{tsang2018}
{Tsang} B.~T.-H.,  {Milosavljevi\'c} M.,  2018, \mn@doi [\mnras]
  {10.1093/mnras/sty1217}, \href
  {http://adsabs.harvard.edu/abs/2018MNRAS.478.4142T} {478, 4142}

\bibitem[\protect\citeauthoryear{{Whitworth}}{{Whitworth}}{1979}]{whitworth1979}
{Whitworth} A.,  1979, \mn@doi [\mnras] {10.1093/mnras/186.1.59}, \href
  {http://adsabs.harvard.edu/abs/1979MNRAS.186...59W} {186, 59}

\bibitem[\protect\citeauthoryear{{Williams} \& {Cieza}}{{Williams} \&
  {Cieza}}{2011}]{williams2011}
{Williams} J.~P.,  {Cieza} L.~A.,  2011, \mn@doi [\araa]
  {10.1146/annurev-astro-081710-102548}, \href
  {http://adsabs.harvard.edu/abs/2011ARA%26A..49...67W} {49, 67}

\bibitem[\protect\citeauthoryear{{Winter}, {Clarke}, {Rosotti}, {Ih},
  {Facchini}  \& {Haworth}}{{Winter} et~al.}{2018}]{winter2018}
{Winter} A.~J.,  {Clarke} C.~J.,  {Rosotti} G.,  {Ih} J.,  {Facchini} S.,
  {Haworth} T.~J.,  2018, \mn@doi [\mnras] {10.1093/mnras/sty984}, \href
  {http://adsabs.harvard.edu/abs/2018MNRAS.478.2700W} {478, 2700}

\bibitem[\protect\citeauthoryear{{Wolfire}, {Hollenbach}, {McKee}, {Tielens}
  \& {Bakes}}{{Wolfire} et~al.}{1995}]{wolfire1995}
{Wolfire} M.~G.,  {Hollenbach} D.,  {McKee} C.~F.,  {Tielens} A.~G.~G.~M.,
  {Bakes} E.~L.~O.,  1995, \mn@doi [\apj] {10.1086/175510}, \href
  {http://adsabs.harvard.edu/abs/1995ApJ...443..152W} {443, 152}

\bibitem[\protect\citeauthoryear{{Wright}, {Drake}, {Drew}, {Guarcello},
  {Gutermuth}, {Hora}  \& {Kraemer}}{{Wright} et~al.}{2012}]{wright2012}
{Wright} N.~J.,  {Drake} J.~J.,  {Drew} J.~E.,  {Guarcello} M.~G.,  {Gutermuth}
  R.~A.,  {Hora} J.~L.,   {Kraemer} K.~E.,  2012, \mn@doi [\apjl]
  {10.1088/2041-8205/746/2/L21}, \href
  {http://adsabs.harvard.edu/abs/2012ApJ...746L..21W} {746, L21}

\makeatother
\end{thebibliography}






\bsp	
\label{lastpage}
\end{document}